\documentclass[twoside]{article}

\usepackage[utf8]{inputenc}
\usepackage{amsfonts}
\usepackage{amsmath,amssymb}
\usepackage{nccmath}
\usepackage{mathrsfs}
\usepackage{graphicx}
\usepackage{subfigure}
\usepackage{tabularx}
\usepackage[all]{xy}
\usepackage{color}
\usepackage{multirow}
\usepackage{booktabs}
\usepackage{enumitem}
\usepackage{hyperref}
\usepackage{natbib}
\usepackage{algpseudocode}
\usepackage{algorithm}
\usepackage{ifthen}
\usepackage{calc}

\newtheorem{proposition}{Proposition}
\newtheorem{lemma}[proposition]{Lemma}
\newtheorem{theorem}[proposition]{Theorem}

\newtheorem{remark}[proposition]{Remark}

\newenvironment{proof}[1][Proof\ ]{\medskip\noindent{\bf #1}\ }{%
\hfill $\Box$\par\quad\par}

\usepackage{times}
\usepackage{enumitem}
\usepackage{multirow}
\usepackage{setspace} 

\def\mcl#1{\mathcal{#1}}

\def\hil{\mcl{H}}
\def\nn{\nonumber}
\def\opn{\operatorname}
\def\mr{\mathrm}

\def\mat{\mathbb{C}^{m\times m}}

\def\uu{u}

\def\eqref#1{(\ref{#1})}

\def\mcl#1{\mathcal{#1}}
\def\blacket#1{\left\langle #1\right\rangle}
\def\blackets#1{\langle #1\rangle}
\def\blacketg#1{\bigg\langle #1\bigg\rangle}

\def\hil{\mcl{H}}

\def\nn{\nonumber}
\def\opn{\operatorname}

\def\mat{\mathbb{C}^{N\times N}}

\def\bu{\mathbf{u}}

\def\bv{\mathbf{v}}

\newenvironment{mythm}[1][]{\medskip\par\noindent{\bfseries #1}\ \,\,\em}{\medskip\par}

\DeclareSymbolFont{EulerExtension}{U}{euex}{m}{n}
\DeclareMathSymbol{\euintop}{\mathop} {EulerExtension}{"52}
\DeclareMathSymbol{\euointop}{\mathop} {EulerExtension}{"48}

\title{Operator-theoretic Analysis of Mutual Interactions\\ in Synchronized Dynamics}

\author{Yuka Hashimoto$^{1,2}$, Masahiro Ikeda$^{2,4}$, Hiroya Nakao$^3$, and Yoshinobu Kawahara$^{2,4}$\\
{\normalsize 1. NTT Corporation}\quad
{\normalsize 2. Osaka University}\\
{\normalsize 3. Institute of Science Tokyo}\quad
{\normalsize 4. Center for Advanced Intelligence Project, RIKEN}}

\date{}

\begin{document}
\twocolumn[
\maketitle
]

\begin{abstract}
Analyzing synchronized nonlinear oscillators is one of the most important and attractive topics in nonlinear science. By understanding the interactions between the oscillators, we can figure out the synchronization process. A promising approach to the analysis of interacting oscillators in nonlinear science is the application of the phase model. In this paper, we propose a data-driven approach to extract mutual interactions of synchronized oscillators based on the phase model. Recently, applying machine learning techniques to estimate models in physics has been actively investigated. We propose an operator-theoretic approach to estimate the phase model of interacting oscillators. We reduce the estimation problem to a multiparameter eigenvalue problem of the Koopman operator, a linear operator that describes a dynamical system. By reducing the problem to a linear algebraic problem, we can theoretically show that the proposed approach is stable with respect to perturbations in the given data.
\end{abstract}

\section{Introduction}
\begin{figure}[t]
    \centering
    \includegraphics[scale=0.35]{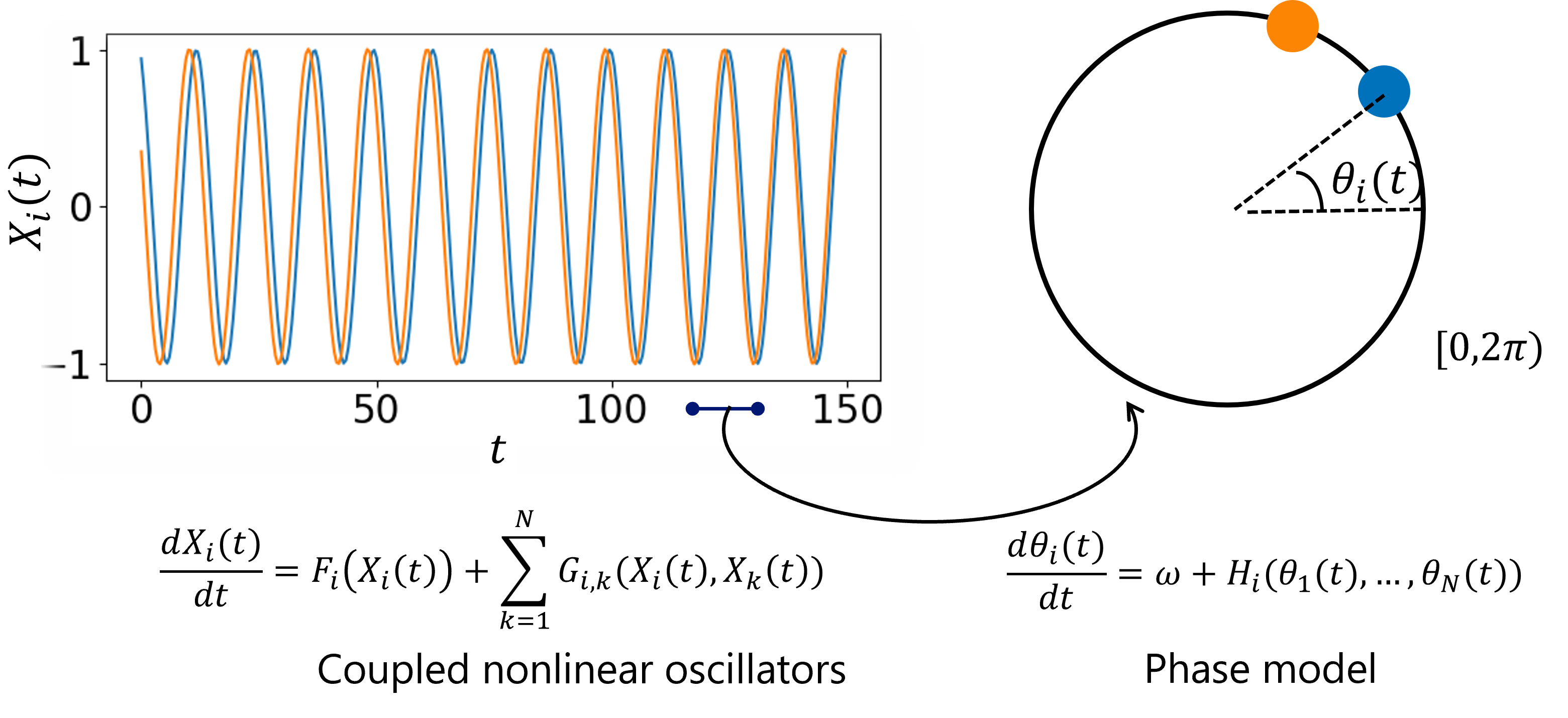}
    \caption{Phase reduction of {coupled nonlinear oscillators} to the phase model.}
    \label{fig:phase_model}
\end{figure}

Analyzing synchronized nonlinear oscillators is important for understanding various phenomena in nature and human society, including energy metabolism, heartbeat, circadian rhythms, and power grids~\citep{winfree01,pikovsky01,sajadi22}.
By understanding the interactions of the oscillators, we can figure out the synchronization process.
An important model-based method in nonlinear science for analyzing synchronized oscillators is the phase reduction, which provides a low-dimensional phase model of interacting oscillators by representing the dynamics of individual oscillators using only their phase values~\citep{kuramoto84,brown04,acebron05,nakao14,nakao16,watanabe19,Pietras19}.
The phase model is represented by intrinsic frequencies, phase functions, and phase coupling functions.
The intrinsic frequency and phase function characterize the dynamics of individual oscillators, and the phase coupling function characterizes the interactions between the oscillators.
Figure~\ref{fig:phase_model} schematically describes the phase reduction.

Although model-driven approaches gives us solid theoretical understandings of the phenomena, we need to know the mathematical models underlying the phenomena, which are not always available.
In addition, even if we know the underlying models, numerical schemes for solving them are sometimes computationally expensive or unstable.
Recently, applying machine learning approaches to estimate models from data has been investigated.
For example, we can use neural ODEs to learn dynamics from time-series data~\citep{chen18,teshima20}.
In addition, we can apply operator learning to obtain nonlinear operators that map functions, such as initial conditions and model parameters, to solutions of the underlying model of data~\citep{kovachki23,li21fourier}.

Operator-theoretic approaches using Koopman operators are attracting much attention as an important tool for analyzing dynamical systems~\citep{koopman31}.
The Koopman operator gives a lifted representation of nonlinear dynamical systems.
Since the Koopman operator is linear even if the original system is nonlinear, 
we can use linear algebra to analyze the dynamics.
An attractive feature of Koopman operators is that we can estimate them from given data and apply them to data-driven problems~\citep{rowley09,kutz13,mezic12,williams15,takeishi17,mezic17,fujii19,klus20,hashimoto19,hu20}.
They enable us to understand the dynamics {using} given time-series data without knowing the underlying dynamical system that generates the data.
This feature makes the operator-theoretic approaches attractive for the machine learning community~\citep{kawahara16,ishikawa18,azencot20,wang23_koopman,liu23}.

In this paper, we propose an operator-theoretic approach, called the Koopman generalized multiparameter eigenvalue (KGME) approach, to estimate the phase model from data to understand the interaction of the synchronizing oscillators.
The {relationship} between Koopman operator theory and phase models has been investigated~\citep{shirasaka17,mauroy12,mauroy13}.
For example, the intrinsic frequency and phase function of a nonlinear oscillator are obtained from the eigenvalue corresponding to the fundamental frequency and associated eigenfunction of the Koopman operator, respectively.
Therefore, by estimating the Koopman operator and computing its eigenvalue and eigenfunction, we can reconstruct the phase model only with given data.
However, most existing studies have focused on estimating the phase model of {\em a single isolated oscillator} from time-series data.
No study has shown that Koopman operator theory can be used to estimate the phase model describing mutual interactions between {\em multiple oscillators}. 

We show that the problem of estimating the phase model {\em for multiple oscillators} is reduced to a generalized multiparameter eigenvalue (GME) problem with the Koopman operator.
This generalizes the existing relationship between the standard eigenvalue problem of the Koopman operator and the phase model {\em for a single isolated oscillator}.
Here, the GME problem is a generalization of the multiparameter eigenvalue problem, which is a generalization of the standard eigenvalue problem and was developed originally for solving ordinary differential equations with multiple parameters~\citep{atkinson11,atkinson72}.
The GME problem is formulated using the product of functions.
Then, to solve the GME problem, we use multiple eigenvectors of the Koopman operator to extract information about the high-frequency components describing the interactions.
These strategies make the algorithm stable with respect to the perturbation of the given data.
Here, as a criterion of stability, we focus on the difference in the resulting learnable parameters caused by the perturbation of the data.
This criterion describes how much the output model is changed due to the input perturbation.
Estimation of the phase model for multiple oscillators without resorting to Koopman operator theory has also been investigated. 
In existing methods, the phase model is estimated by, for example, fitting the data to the Fourier series~\cite{Blaha11,tokuda19}.
However, in this case, the change in the output model due to the input perturbation depends on $M^2\sqrt{M}$, where $M$ is the number of the frequency components.
Our formulation of the problem using the product of functions alleviates the dependency.
As a naive method with the Koopman operator, we can also use the powers of a single eigenvector of the Koopman operator to extract the high-frequency components.
In this case, the change in the output model depends linearly on the number of frequency components.
On the other hand, for the proposed KGME approach, the change in the output model does not depend on the number of frequency components because of the effect of using the multiple eigenvectors of the Koopman operator.
The results of the stability are summarized in Table~\ref{tab:summary}. 

Our main contributions are summarized as follows:\vspace{-.2cm}
\begin{itemize}[nosep,leftmargin=*]
\item We propose the KGME approach to estimate the phase model describing mutual interactions between multiple oscillators by reducing the problem to a generalized multiparameter eigenvalue problem of the Koopman operator.
\item We theoretically show that the proposed approach is stable with respect to the perturbation in the given data compared to the existing method with the Fourier series and a naive method with the Koopman operator.
\end{itemize}
Our study expands the applicability of Koopman operators to the analysis of synchronization and interactions of oscillators, and sheds light on
the connection between operator-theoretic data analysis and nonlinear science.

\newcolumntype{C}{>{\centering\arraybackslash}X}
\begin{table}[t]
\caption{Comparison of the dependency of the change in the output model on the number of frequency components $M$ in the model for estimating the phase model.}\label{tab:summary}
\small
\centering
\begin{tabularx}{0.4\textwidth}{c|c}
\hline
     Approach & Dependency\\
     \hline
 Fourier (existing approach)   &  $M^2\sqrt{M}$\\
 Naive approach with Koopman & $M$ \\
 {\bf KGMP (proposed approach)} & {\bf 1}\\
\hline
\end{tabularx}
\end{table}

\section{Preliminaries}
We review two important notions; phase reduction and Koopman operator.
\subsection{Phase reduction of weakly interacting oscillators}
Let $\mcl{X}=\mathbb{R}^d$ and consider the following dynamical system:
\begin{equation}
\begin{aligned}
\frac{dX(t)}{dt}=F(X(t)),
\end{aligned}\label{eq:DS}
\end{equation}
where $X(t)\in\mcl{X}$ and $F:\mcl{X}\to\mcl{X}$ is a nonlinear map.
Let $\tilde{\Phi}:[0,\infty)\times \mcl{X}$ be the flow map of the dynamical system~\eqref{eq:DS}, i.e., $\tilde{\Phi}(\cdot,x)$ is the trajectory starting from the initial point $x$.
Analyzing dynamical systems that behave as oscillators has been one of the major topics in nonlinear science.
We focus on the case where the dynamical system $F$ has a limit cycle attractor, i.e., the trajectory tends to be a stationary periodic oscillation after a sufficiently long time.
More precisely, we assume that there exists a periodic trajectory $Y:[0,\infty)\to \mcl{X}$ that satisfies $\frac{dY(t)}{dt}=F(Y(t))$ and $Y(t)=Y(t+T)$ for some $T>0$ and any $t\in [0,\infty)$, and there exists an open set $\Omega\subseteq \mcl{X}$ such that $\bigcap_{t\ge 0} \tilde{\Phi}(t,\Omega)=\{Y(t)\,\mid\,t\in [0,\infty)\}$.

When multiple oscillators interact, they can exhibit mutual synchronization.
In this case, we consider the $N$ oscillators $X_1(t),\ldots,X_N(t)\in\mcl{X}$ governed by $F$ and describe the interaction between the $i$th oscillator and the $k$th oscillator using a function $G_{i,k}:\mcl{X}\times \mcl{X}\to\mcl{X}$.
The dynamical system in this case is 
\begin{equation}
\begin{aligned}
\frac{dX_i(t)}{dt}=F(X_i(t))+&\sum_{k=1}^{N}G_{i,k}(X_i(t),X_k(t))
\end{aligned}\label{eq:NWDS}
\end{equation}
for $i=1,\ldots,N$. 
Let ${\Phi}:[0,\infty)\times \mcl{X}^N$ be the flow map of the dynamical system~\eqref{eq:NWDS}.
We focus on the case where the oscillators $X_1,\ldots,X_N$ are dominated by a common intrinsic frequency $\omega$, interact weakly with each other, and synchronize after a sufficiently long time ~\cite{nakao16,Pietras19}. 
More precisely, we assume $G_{i,j}(x,x)=0$, $G_{i,k}$ is sufficiently small, $G_{i,k}(X_i(t),X_k(t))\to 0$ as $t$ goes to infinity, and $\Vert X_i(t)-Y(t)\Vert\to 0$ as $t$ goes to infinity for any $i=1,\ldots,N$.
This type of coupled dynamical model describes the synchronization phenomena, which are important in biological and engineered systems~\citep{winfree01,pikovsky01,sajadi22}.

In this case, we can construct a map $\Theta:\Omega\to [0,2\pi)$ that maps the physical state $X_i(t)$ to the corresponding phase $\theta_i(t)$, and the dynamics can be approximately reduced to the following phase model:
\begin{equation}
\begin{aligned}
\frac{d\theta_{i}(t)}{dt}=\omega+&\Gamma_{i}(\theta_{i}(t)-\theta_{1}(t),\ldots,\theta_{i}(t)-\theta_{N}(t)) 
\end{aligned}
\label{eq:phase_model_couple}
\end{equation}
for $i=1,\ldots,N$.
Here, 
$\Gamma_i$ is the phase coupling function that describes the effect of the phase differences between the oscillators exerted on the element $i$~\cite{nakao16}.
Note that since the interactions are assumed to be weak, we can approximate the effect only with the phase differences rather than the absolute phase values~\cite{kuramoto84}.

The phase model enables us to perform a detailed analysis of the synchronization due to interactions among oscillators.


\subsection{Koopman operator}\label{subsec:koopman}
While the maps $F$ and $G_{i,k}$ are in general nonlinear, the Koopman operator is a linear operator that gives us a tool for analyzing them with linear algebraic approaches.
This perspective was originally introduced by \citet{koopman31} and later extended to dissipative systems by \citet{mezic05}. 

Let $\tilde{\hil}$ be a Hilbert space of $\mathbb{C}$-valued functions on $\Omega$.
Consider the Koopman operator $\tilde{K}^t$ of the dynamical system $F$, defined as 
\begin{align*}
\tilde{K}^tv(x)=v(\tilde{\Phi}(t,x))    
\end{align*}
for $v\in\tilde{\hil}$, $t\in [0,\infty)$, and $x\in\Omega$.
We also consider the Koopman operator of the coupled dynamical system~\eqref{eq:NWDS} with interactions.
Since there are $N$ dynamical elements, we consider the Koopman operator on a $N$-dimensional vector-valued function space.
Let $\hil$ be a Hilbert space of $\mathbb{C}^N$-valued functions on $\Omega^N$.
Consider the Koopman operator ${K}^t$ of the dynamical system~\eqref{eq:NWDS}, defined as 
\begin{align*}
{K}^tv(x)=v({\Phi}(t,x))    
\end{align*}
for $v\in{\hil}$, $t\in [0,\infty)$, and $x\in\Omega^N$.
For any $\tilde{v}\in\tilde{\hil}$, we can construct a map $v:\Omega^N\to\mathbb{C}^N$ by $v(x_1,\ldots,x_N)=[\tilde{v}(x_1),\ldots,\tilde{v}(x_N)]$.
We assume that for any $\tilde{v}\in\tilde{\hil}$, $v$ is contained in $\hil$.
In this paper, as the Hilbert spaces $\tilde{\hil}$ and $\hil$, we use a reproducing kernel Hilbert space (RKHS)~\cite{scholkopf01} and vector-valued reproducing kernel Hilbert space (vvRKHS)~\cite{kadri16}, respectively, which enables us to theoretically analyze the proposed approach using the reproducing property. 
For a given RKHS, we can easily construct a vvRKHS that satisfies the above assumption (See appendix~\ref{ap:vvrkhs} for more details).

An advantage of applying Koopman operators is that we can estimate them only with given data, which has been actively investigated~\citep{williams15,kawahara16,mezic17,hashimoto19}.
Let $x_0,\ldots,x_T\in\Omega^N$ be given data generated by the dynamical system~\eqref{eq:NWDS} at time $t_0,\ldots,t_T$ with a constant time interval $\Delta t$, i.e., $t_{i+1}=t_i+\Delta t$.
We assume that $L$ sequences $\{x^1_0,\ldots,x^1_T\},\ldots,\{x^L_0,\ldots,x^L_T\}$ of data are given.
We denote the set of these sequences by $S$.
We generate finite-dimensional subspaces of $\tilde{\hil}$ and $\hil$ using the given data $S$ and estimate the Koopman operators $\tilde{K}^t$ and $K^t$ on the subspaces.
The estimation of the Koopman operator in vvRKHS is detailed in Appendix~\ref{ap:Koopman_estimation}.
They are obtained by using an operator $Q_S:\mathbb{C}^n\to\hil$ constructed with orthonormal basis of the finite-dimensional subspace.
Here, $n$ is the dimension of the subspace.
The estimation of the Koopman operator $K^t$ is obtained as $Q_SQ_S^*K^tQ_SQ_S^*$, where $^*$ represents the adjoint, and the representation matrix of this operator is $Q_S^*K^tQ_S$, which is denoted by $\mathbf{K}^t_S$.
See Appendix~\ref{sec:Koopman_estimation} for more details.
In the following, to simplify the notation, we denote by $K$ and $\tilde{K}$ the Koopman operators $K^{\Delta t}$ and $\tilde{K}^{\Delta t}$ for the time interval $\Delta t$, respectively.

\section{KGME approach for estimating phase models}
We propose a KGME approach for estimating phase models.
We first estimate the phase function $\Theta$ and the intrinsic frequency $\omega$.
Then, we estimate the phase coupling function $\Gamma_{i}$ based on the Fourier series.
These two processes are archived via linear algebraic approaches using the Koopman operators defined in Subsection~\ref{subsec:koopman}.
As we will see in Section~\ref{sec:stability}, thanks to the use of linear algebra, the proposed approach is stable with respect to the perturbation of given data.

\subsection{Estimation of the phase function}\label{subsec:phase_function}
We first estimate the phase function $\Theta$ and the intrinsic frequency $\omega$ via the Koopman operator $\tilde{K}$ of $F$.
As proposed by~\citet{shirasaka17,mauroy18,mauroy13}, the phase function $\Theta$ can be obtained by computing an eigenvector of the Koopman operator $\tilde{K}$ corresponding to the frequency $\omega$.
In the existing frameworks with Koopman operators, they focus on the oscillators without interactions.
On the other hand, in our case, we focus on a more complicated and realistic situation where multiple oscillators interact.
For this purpose, we consider the $M$ eigenvectors of the Koopman operator corresponding to the frequency $j\omega$ for $j\in\mathbb{N}$.

Let $\lambda\in\mathbb{C}$ be an eigenvalue of the Koopman operator $\tilde{K}$ with $\vert\lambda\vert=1$ and let $u\in\tilde{\hil}$ be the corresponding eigenvector, that is,
\begin{align*}
\tilde{K}u=\lambda u.
\end{align*}
Let $\omega\in[0,2\pi)$ that satisfies $\lambda=\mr{e}^{\sqrt{-1}\omega\Delta t}$.
By the definition of $\tilde{K}$, for $i=1,\ldots,N$, we have
\begin{align}
   u(X_i(t+\Delta t))= \mr{e}^{\sqrt{-1}\omega\Delta t}u(X_i(t)).\label{eq:standard_eig}
\end{align}
Let $\theta_{i}(t)=\opn{arg}(u(X_i(t)))$.
By Eq.~\eqref{eq:standard_eig}, the identity $\vert u(X_i(t+\Delta t))\vert=\vert u(X_i(t))\vert$ holds. 
Thus, we have
\begin{align*}
\theta_{i}(t+\Delta t)=\omega\Delta t+\theta_{i}(t),
\end{align*}
which means that each element $i$ oscillates with frequency $\omega$.
The map $\opn{arg}(u(\cdot))$ transforms $X_i(t)$ into the corresponding phase $\theta_i(t)$.
Thus, it is equal to the phase function $\Theta$ corresponding to the intrinsic frequency $\omega$.

An advantage of applying Koopman operators is that we can also obtain the phase function corresponding to $j\omega$ for $j\in\mathbb{N}$ from the eigenvector corresponding to the eigenvalue $\mr{e}^{\sqrt{-1}j\omega\Delta t}$.
The following lemma is derived directly from Eq.~\eqref{eq:standard_eig}.
\begin{lemma}
Let $u\in\tilde{\hil}$ be the eigenvector of $\tilde{K}$ corresponding to the eigenvalue $\mr{e}^{\sqrt{-1}\omega \Delta t}$.
Then, if $u^j\in\tilde{\hil}$, then it is the eigenvector corresponding to the eigenvalue $\mr{e}^{\sqrt{-1}j\omega \Delta t}$ for $j\in\mathbb{Z}$.
Here, $u^j(x)=u(x)^j$.
\end{lemma}

\begin{remark}
Once we obtain the single eigenvector $u$ corresponding to the eigenvalue $\mr{e}^{\sqrt{-1}\omega\Delta t}$, we can also obtain the eigenvector corresponding to the eigenvalue $\mr{e}^{\sqrt{-1}j\omega\Delta t}$ by computing the power of $u$. 
Alternatively, we can also obtain the phase function corresponding to the frequency $j\omega$ by computing $j\Theta$.
However, both of these computations make the algorithm for estimating the phase coupling function unstable.
We will investigate the stability of the algorithm for estimating the phase coupling function in Section~\ref{sec:stability}.
\end{remark}

\subsection{Estimation of the phase coupling function}\label{subsec:phase_coupling_functions}
To obtain the phase coupling function $\Gamma_i$, we use the following model with the Fourier series truncated to $M$ terms:
\begin{align}
&\Gamma_{i}(\theta_i(t)-\theta_1(t),\ldots,\theta_i(t)-\theta_N(t))\nn\\
&\ = \frac{2}{M(M+1)}\sum_{j=1}^M\opn{arg}\bigg(\sum_{k=1}^N{a}_{i,k}^j e^{\sqrt{-1}j(\theta_{k}(t)-\theta_{i}(t))}\bigg),\label{eq:gamma_fourier}
\end{align}
where $a_{i,k}^j\in\mathbb{C}$ for $i,k=1,\ldots,N$ and $j=1,\ldots,M$ is the learnable parameter.
Applying the Fourier series to estimate the phase coupling function has been proposed~\citep{tokuda19}.
In this paper, we use the Koopman operators to train the model~\eqref{eq:gamma_fourier}, which makes the algorithm more stable.

On the other hand, consider the problem with the Koopman operator of finding $\{a_{i,k}^j\}_{i,k,j}$ satisfying
\begin{align}
 &\mr{e}^{-\sqrt{-1}\omega\Delta t}K\uu^1\odot\cdots\odot \mr{e}^{-\sqrt{-1}M\omega\Delta t} K\uu^M\nn\\
 &=\bigg(\sum_{i,k=1}^Na^1_{i,k}B_{i,k}\uu^1\bigg)\odot\cdots\odot \bigg(\sum_{i,k=1}^Na^M_{i,k}B_{i,k}\uu^M\bigg),\label{eq:multiparam_eig_prod}   
\end{align}
where we abuse the notation and denote by $\uu^j$ the map $\Omega^N\ni [x_1,\ldots,x_N]\mapsto [u(x_1)^j,\ldots,u(x_N)^j]\in\mathbb{C}^N$, and $B_{i,k}$ is a linear operator on $\hil$ that maps $[v_1,\ldots,v_k]\in\hil$ to the vector in $\hil$ whose $i$th element is $v_k$ and the other elements are $0$.
In addition, $\mathbf{v}\odot \mathbf{w}$ is defined as the elementwise product, i.e., $\bv\odot\mathbf{w}=[v_1w_1,\ldots,v_Nw_N]$.

The following theorem shows that we can reduce the problem of the estimation of the model~\eqref{eq:gamma_fourier} to the problem~\eqref{eq:multiparam_eig_prod}.
In the following, the proofs of the statements are documented in Appendix~\ref{ap:proofs}.

\begin{theorem}\label{thm:gme_phase_model}
The solution $\{a_{i,k}^j\}$ of the problem~\eqref{eq:multiparam_eig_prod} satisfies the discretized version of the phase model~\eqref{eq:phase_model_couple},
\begin{align}
&\theta_i(t+\Delta t)=\omega \Delta t +\theta_i(t)\nn\\
&\qquad+{\Delta t}\Gamma_i(\theta_i(t)-\theta_1(t),\ldots,\theta_i(t)-\theta_N(t))\label{eq:phase_coupling_discretized}
\end{align}
with $\Gamma_i$ in the form of Eq.~\eqref{eq:gamma_fourier}.
\end{theorem}

Note that $a_{i,k}^j$ for $j=1,\ldots,M$ describes the effect of element $k$ on element $i$ in the interaction.
In addition, $M$ is the parameter for the complexity of $\Gamma_i$ and determines the number of Fourier components in $\Gamma_i$. We can describe higher frequency components as $M$ becomes larger.

\begin{remark}
The problem~\eqref{eq:multiparam_eig_prod} is a generalization of the multiparameter eigenvalue problem.
The multiparameter eigenvalue problem is formulated as follows: given matrices $A_1,\ldots,A_n$, find a vector $v$ and values $\lambda_1,\ldots,\lambda_n\in\mathbb{C}$ that satisfy
\begin{align*}
\sum_{i=1}^n\lambda_iA_iu=0.
\end{align*}
Indeed, in the case of $M=1$, by setting $n=N^2+1$, $\lambda_1=\mr{e}^{\sqrt{-1}\omega\Delta t}$, $A_1=K$, $\lambda_{(i-1)N+k+1}=a^1_{i,k}$, $A_{(i-1)N+k+1}=B_{i,k}$ for $i,j=1,\ldots,N$, and $v=u^1$, we can regard Eq.~\eqref{eq:multiparam_eig_prod} as a multiparameter eigenvalue problem.
\end{remark}

\subsection{Practical computations}\label{subsec:practical}
\paragraph{Problem formulation}
The Koopman operators appearing in Subsections~\ref{subsec:phase_function} and \ref{subsec:phase_coupling_functions} are operators on infinite-dimensional spaces, and without knowing the dynamical systems Eqs.~\eqref{eq:DS} and \eqref{eq:NWDS}, we do not have the real Koopman operators.
Thus, we need to estimate them using given data.
As we discussed in Subsection~\ref{subsec:koopman}, given time-series data ${S}=\{x_0,\ldots,x_T\}\subset\Omega$, we can estimate the Koopman operators underlying ${S}$.
As stated in Subsection~\ref{subsec:koopman}, the estimated operator of $K$ is written as $Q_{{S}}\mathbf{K}_{{S}}Q_{{S}}$.
In practical computations, we first estimate $\tilde{K}$ using time-series data $S_1:=\{x_{T_1},\ldots,x_T\}$ with sufficiently large $T_1$, and obtain the matrix $\mathbf{\tilde{K}}_{S_1}$.
This is because after a sufficiently long time, the oscillators tend to synchronize with each other completely and the interaction $G_{i,k}$ becomes closer to $0$.
Instead of $\tilde{K}$ in Subsection~\ref{subsec:phase_function}, we compute the eigenvalues and eigenvectors of $\mathbf{\tilde{K}}_{S_1}$.
Then we obtain the eigenvalues $\mr{e}^{\sqrt{-1}j\omega_{S_1}\Delta t}$ and the corresponding eigenvectors $\bu_{S_1}^j$ ($j=1,\ldots,M)$.
We note that $Q_{S_1}\bu_{S_1}^j$, which is an estimation of $\uu^j$, is obtained as an eigenvector of the estimated Koopman operator, not by computing the power of $Q_{S_1}\bu_{S_1}^1$.
We then use $S_2:=\{x_{0},\ldots,x_{T_2}\}$ with some $T_2>0$, the time-series before the synchronization, for estimating $K$, and obtain $\mathbf{K}_{S_2}$.
By replacing $K$ with $Q_{S_2}\mathbf{K}_{S_2}Q_{S_2}^*$ and $u^j$ with $Q_{S_1}\bu_{S_1}^j$ in Eq.~\eqref{eq:multiparam_eig_prod}, we obtain
\begin{align}
 &\mr{e}^{-\sqrt{-1}\omega_{S_1}\Delta t}\mathbf{K}_{S_2}P\mathbf{u}_{S_1}^1\odot\cdots\odot \mr{e}^{-\sqrt{-1}M\omega_{S_1}\Delta t} \mathbf{K}_{S_2}P\bu_{S_1}^M\nn\\
 &=\bigg(\sum_{i,k=1}^Na^1_{i,k}PB_{i,k}\mathbf{u}_{S_1}^1\bigg)\odot\cdots\odot \bigg(\sum_{i,k=1}^Na^M_{i,k}PB_{i,k}\mathbf{u}_{S_1}^M\bigg),\label{eq:multiparameter_estimated}
\end{align}
where $P=Q_{S_2}^*Q_{S_1}$.

\paragraph{Optimization}
To find an optimal $\{a_{i,k}^j\}_{i,k,j}$ in Eq.~\eqref{eq:multiparameter_estimated}, we set a loss function and minimize it, for example, using a gradient method.
If the interaction $G_{i,k}$ is zero, then we have
\begin{align}
\mr{e}^{-\sqrt{-1}j\omega\Delta t}K\uu^j=\sum_{i,k=1}^Na^1_{i,k}B_{i,k}\uu^j\label{eq:nointeraction_eig}
\end{align}
with $[a_{i,k}^j]_{i,k}=I$ for $j=1,\ldots,M$.
Here, $I$ is the identity matrix.
Since $G_{i,k}$ is small, we expect that Eq.~\eqref{eq:nointeraction_eig} holds approximately even for the case of $G_{i,k}\neq 0$.
Therefore, we set the following loss function:
\begin{align}
L^{\opn{KGME}}(S)=\sum_{j=1}^M\Vert L^{\opn{KGME}}_{j}(S)\Vert^2,\label{eq:loss}
\end{align}
where
\begin{align*}
&L^{\opn{KGME}}_{j}(S)=\mathbf{K}_{S_2}P\mathbf{u}_{S_1}^j-\lambda_{S_1,j}\sum_{i,k=1}^Na^j_{i,k}PB_{i,k}\mathbf{u}_{S_1}^j,
\end{align*}
where $\lambda_{S_1,j}$ is the eigenvalue of $\mathbf{\tilde{K}}_{S_1}$ that is the closest to $\mr{e}^{\sqrt{-1}j\omega_{S_1}\Delta t}$.

\section{Stability analysis of the KGME approach}\label{sec:stability}
The proposed KGME approach has the advantage of algorithmic stability.
We focus on the case where we use the gradient descent method to obtain an optimal $\{a_{i,k}^j\}$.
The updated parameter $a^{j,\opn{new}}_{i,k}$ with one iteration of the gradient descent method for a loss function $L$ is written as
\begin{align*}
a^{\opn{new},j}_{i,k}=a^j_{i,k}-\eta(\nabla L(S))_{i,k},
\end{align*}
where $\nabla$ is the gradient with respect to $(a_{i,k}^j)_{i,j,k}$ and $\eta>0$ is the learning rate.
By deriving a bound of the difference $\vert \nabla L(S)-\nabla L(S')\vert$ between gradients, we obtain the following theorem regarding the stability of the KGME approach with respect to the perturbation of the time-series from $S$ to $S'$.
Let $\lambda_{S_1,i}$ be eigenvectors of $\mathbf{\tilde{K}}_{S_1}$.
Let $U_{S_1}$ be the matrix whose columns are right eigenvectors of $\mathbf{\tilde{K}}_{S_1}$ and $W_{S_1}$ be the matrix whose columns are left eigenvectors of $\mathbf{\tilde{K}}_{S_1}$ and satisfy $W_{S_1}^*U_{S_1}=I$.
In addition, let $a_{S,i,k}^j$ be the learnable parameter trained with $S$, $A_{S}^j$ be the $N$ by $N$ matrix whose $(i,k)$-entry is $a_{S,i,k}^j$, and $\alpha_S=\max_{i\neq j}{1}/(\lambda_{S_1,i}-\lambda_{S_1,j})$.
\begin{theorem}\label{thm:stability}
Assume 
$\Vert Q_{S_1}-Q_{S_1'}\Vert\le \epsilon/(2\Vert\tilde{K}\Vert)$, $\Vert Q_{S_2}-Q_{S_2'}\Vert\le \epsilon/(2\Vert {K}\Vert)$ and $\vert a^j_{S,i,k}-a^j_{S',i,k}\vert\le\epsilon$ for some $\epsilon>0$.
Assume, in addition, for any time-series $S$, $\Vert A_S^j\Vert\le D_A$, $\Vert U_S\Vert\le D_U$, and $\Vert W_S\Vert\le D_W$ for some $D_A,D_U,D_W>0$.
Then, we have
\begin{align*}
&\vert a^{\opn{new},j}_{S,i,k}-a^{\opn{new},j}_{S',i,k}\vert \\
&\le \epsilon+2\epsilon\eta N  D_U^2 (1+N+D_UD_WD_A\\
&\quad+2(\Vert K\Vert+D_A)D_UD_W\alpha_S)+O(\epsilon^2).
\end{align*}
\end{theorem}

Note that $Q_{S_1}$ characterizes the subspace obtained from the given time-series $S_1$ to approximate the Koopman operator.
Thus, Theorem~\ref{thm:stability} shows that if the approximation spaces generated by $S$ and $S'$ are sufficiently close, then with one step of the gradient descent method, the difference in the learnable parameter $a_{i,k}^j$ expands with the rate $1+2\eta N  D_U^2 (1+N+D_UD_WD_A+2(\Vert K\Vert+D_A)D_UD_W\alpha_S)$, which does not depend on $j$ and $M$.

To show Theorem~\ref{thm:stability}, we apply the following lemma~\citep{bamieh22}.

\begin{lemma}\label{lem:stability_eig}
Assume $\Vert \mathbf{\tilde{K}}_{S_1}-\mathbf{\tilde{K}}_{S_1'}\Vert\le \epsilon$ for some $\epsilon>0$.
Then, we have
\begin{align*}
\Vert U_{S_1}-U_{S_1'}\Vert 
&\le \epsilon\Vert U_{S_1}\Vert^2\Vert W_{S_1}\Vert\alpha_{S}+O(\epsilon^2)\\
\vert\lambda_{S_1,i}-\lambda_{S_1',i}\vert&\le \epsilon\Vert U_{S_1}\Vert\Vert W_{S_1}\Vert+O(\epsilon^2).
\end{align*}
\end{lemma}

\subsection{Comparison to the approach with computing the power of the eigenvector}\label{subsec:pow}
In the KGME approach, we obtain $\bu_{S_1}^j$ from the eigenvector of $\mathbf{\tilde{K}}_{S_1}$ corresponding to the eigenvalue $\mr{e}^{\sqrt{-1}j\omega\Delta t}$.
On the other hand, naively, we can also obtain $\bu_{S_1}^j$ by considering the power of $Q_{S_1}\bu_{S_1}^1$.
Note that $\bu_{S_1}^1$ is a finite-dimensional vector, and the elementwise power of $\bu_{S_1}^1$ does not give an estimation of the function $\uu^j$.
On the other hand, we can estimate $\uu^j(x)$, the value of $\uu^j$ at $x$, as $(Q_{S_1}\bu_{S_1}^1(x))^j$. 
In this case, the simplest loss function is
\begin{align*}
{L}^{\opn{pow}}(S)=\sum_{j=1}^M\sum_{x\in S_2}\Vert {L}^{\opn{pow}}_{j,x}(S_1)\Vert^2,
\end{align*}
where 
\begin{align*}
&{L}^{\opn{pow}}_{j,x}(S_1)
=(Q_{S_1}\mathbf{u}_{S_1}^1(\Phi(\Delta t,x)))^j\\
&-\mr{e}^{\sqrt{-1}j\omega_{S_1}\Delta t}\sum_{i,k=1}^Na^j_{i,k}B_{i,k}(Q_{S_1}\mathbf{u}_{S_1}^1(x))^j.
\end{align*}
However, in this case, the algorithm is unstable compared to the KGME approach since it computes the power of $Q_{S_1}\bu_{S_1}^1(x)$.
More precisely, we have the following proposition.
\begin{proposition}\label{prop:stability_single_eig}
Let $\hil$ be an RKHS associated with a kernel $k$ that satisfies $\sup_{x\in\Omega}\Vert k(x,x)\Vert \le C$ for some $C>0$.
Let $\opn{Ev}_S:C(\Omega)^{N^2}\to\mathbb{C}^{N^2}$ be defined as $v\mapsto v(x)$, where $C(\Omega)$ is the Banach space of continuous functions on $\Omega$.
Assume $\Vert\mr{Ev}_{S_2}-\mr{Ev}_{S_2'}\Vert\le \epsilon$, $\Vert Q_{S_1}-Q_{S_1'}\Vert\le \epsilon/(2\Vert\tilde{K}\Vert)$, and $\vert a^j_{S,i,k}-a^j_{S',i,k}\vert\le\epsilon$ for some $\epsilon>0$.
Assume for any time-series $S$, $\Vert \opn{Ev}_S\Vert\le D_E$, $\Vert A_S^j\Vert\le D_A$, $\Vert U_S\Vert\le D_U$, $\Vert Q_{S_1}\bu^1(x)^j\Vert\le D_U$, and $\Vert W_S\Vert\le D_W$ for some $D_E,D_A,D_U,D_W>0$.
Then, we have
\begin{align*}
&\vert a^{\opn{new},j}_{S,i,k}-a^{\opn{new},j}_{S',i,k}\vert \\
&\le \epsilon+2\epsilon N  D_U^2(1+D_A+D_E +ND_E)\nn\\
&\quad +2\epsilon N  D_U^2j\bigg(D_A \Delta t\nn\\
&\quad +2D_E\sqrt{N}C^{1/2}(1+D_A)\bigg(D_U^2D_W\alpha_S+\frac{D_U}{2\Vert\tilde{K}\Vert}\bigg)\bigg)\nn\\
&\quad+O(\epsilon^2).
\end{align*}
\end{proposition}
In this case, the difference in the learnable parameter $a^j_{i,k}$ depends linearly on $j$, instead of the independence of $j$ and $M$ for the KGME approach.

\subsection{Comparison to the approach with the Fourier series}\label{subsec:fou}
We can also estimate $\Gamma_{i}$ directly using the Fourier series as
\begin{align}
&\Gamma_{i}(\theta_i(t)-\theta_1(t),\ldots,\theta_i(t)-\theta_N(t))\nn\\
&\ = \sum_{j=1}^M\sum_{k=1}^N{a}_{i,k}^j e^{\sqrt{-1}j(\theta_{k}(t)-\theta_{i}(t))},\label{eq:gamma_fourier_direct}
\end{align}
where $a_{i,k}^j\in\mathbb{C}$ is the learnable parameter.
We can obtain the phase function $\Theta_{S_1}$ and the corresponding intrinsic frequency $\omega_{S_1}$ using the eigenvalue and eigenvector of the estimated Koopman operator $\tilde{\mathbf{K}}_{S_1}$.
We may also use any approach to obtain $\Theta_{S_1}$ and $\omega_{S_1}$ from given time-series data $S_1$ by, for example, the linear interpolation or the Hilbert transform~\cite{kralemann13,stankovski15}.
Then, the simplest loss function for the model~\eqref{eq:gamma_fourier_direct} is 
\begin{align*}
L^{\opn{Fou}}(S)=\sum_{x\in S_2}\Vert L^{\opn{Fou}}_{1,x}(S_1)\Vert^2,
\end{align*}
where the $i$th element of $L^{\opn{Fou}}_{1,x}(S_1)$ is defined as
\begin{align*}
&L^{\opn{Fou}}_{1,x}(S_1)_i
=\Theta_{S_1}(\Phi(\Delta t,x))_i-\omega_{S_1} \Delta t \\
&\qquad- \Theta_{S_1}(x)_i-\sum_{j=1}^M\sum_{k=1}^N{a}_{S,i,k}^j e^{\sqrt{-1}j(\Theta_{S_1}(x)_k-\Theta_{S_1}(x)_i)}.
\end{align*}
The following proposition shows that computing $j\Theta_{S_1}$ and setting the simple loss function make the algorithm unstable. 
\begin{proposition}\label{prop:stability_fou}
Assume there exists $\epsilon>0$ such that for any $x\in\Omega$ and $i,k=1,\ldots,N$, $\vert \Theta_{S_1}(x)-\Theta_{S_1'}(x)\vert_i\le \epsilon$, $\vert\omega_{S_1}-\omega_{S_1'}\vert\le \epsilon$, and $\vert a_{S,i,k}^j-a_{S',i,k}^j\vert\le \epsilon$.
Assume for any time-series $S$, $\Vert \opn{Ev}_S\Vert\le D_E$, $\vert A_S^j\vert\le D_A$.
Then, we have
\begin{align*}
&\vert a^{\opn{new},j}_{S,i,k}-a^{\opn{new},j}_{S',i,k}\vert \\
&\le \epsilon+2\epsilon \eta N\sqrt{M}(2\sqrt{N}+2\pi\Delta t+MD_A\sqrt{N}\\
&\quad +D_E(1+\Delta t+MN\sqrt{N}+2M\sqrt{N}D_Aj))\\
&\quad+2\epsilon\eta D_E N M(2\sqrt{N}+2\pi\Delta t+MD_A\sqrt{N}).
\end{align*}
\end{proposition}
In this case, the difference in the learnable parameter $a_{i,k}^j$ depends on $M\sqrt{M}j$.

\begin{remark}
The formulation~\eqref{eq:multiparam_eig_prod} with the product and setting the loss function as the sum of that for each Fourier series as Eq.~\eqref{eq:loss} makes the dependence of the difference in the learnable parameter $a_{i,k}^j$ on $j$ less than or equal to linear.
Furthermore, using multiple eigenvectors instead of computing the power of an eigenvector makes the difference in the learnable parameter independent of $j$ and $M$.
This is a main advantage of applying the KGME approach.
\end{remark}

\section{Numerical results}
In this section, we numerically confirm the theoretical investigations in Section~\ref{sec:stability} and show the validity of the proposed KGME approach.
We consider two important models for the physical and biological sciences: the van der Pol model and FitzHugh--Nagmo model.
For example, they are used in understanding the dynamics of spiking neurons.
In addition, we apply the KGME approach to real-world data.

\subsection{van der Pol model}\label{subsec:vdp}
As the first example, we consider the van der Pol model with $N=2$ and $\Omega=(-2.5,2.5)\times (-2.5,2.5)\subseteq \mathbb{R}^2$, which is described by Eq.~\eqref{eq:NWDS} with $F(x)=[x_2,\mu(1-x_1^2)x_2-x_1]$ and $G_{i,k}(x,y)=[0,-0.025(x_2-y_2)]$ for $x=[x_1,x_2],y=[y_1,y_2]\in\Omega$.
We set $\mu=0.3$.
We generated a time-series $S=\{x_0,\ldots,x_T\}\subseteq\Omega^2$ with $T=10000$ and $\Delta t=0.001$ based on Eq.~\eqref{eq:NWDS}.
Then, we set $T_1=T_2=9000$ and constructed estimations $\mathbf{\tilde{K}}_{S_1}$ and $\mathbf{K}_{S_2}$ of the Koopman operators $\tilde{K}$ and $K$.
We also generated five different perturbed time-series $S'$ by perturbing the initial value $x_0$ of $S$ to $x_0+\xi$, where $\xi$ is randomly drawn from the normal distribution with mean 0 and standard deviation 0.0001.
For each $S'$ and each loss function $L$ defined in Subsections~\ref{subsec:practical} (KGME), \ref{subsec:pow} (called ``Power''), and \ref{subsec:fou} (called ``Fourier'')., we computed the sensitivity $\Vert \nabla L(S)-\nabla L(S')\Vert/\Vert \nabla L(S)\Vert$ of the gradient of the loss function $L$.
Regarding the initial value and the regularization for optimizing the learnable parameter $a^j_{i,k}$, see Appendix~\ref{ap:experiment}.
Figure~\ref{fig:graddiff} shows the results.
We can see that with the KGME approach, the sensitivity of the gradient is the smallest and does not increase even though $M$ increases.
On the other hand, with the Power and Fourier approaches, the sensitivity of the gradient increases monotonically as $M$ grows. 
This result numerically supports the theoretical results in Section~\ref{sec:stability}.

\begin{figure}[t]
    \centering
    \includegraphics[scale=0.6]{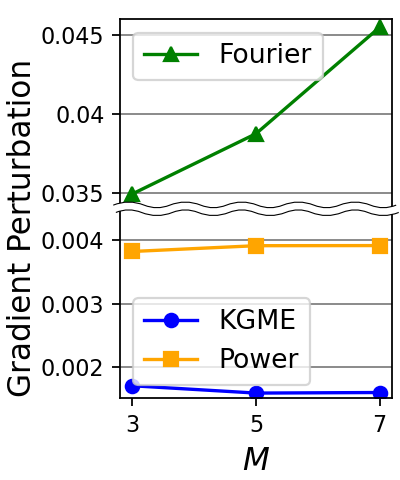}\vspace{-.5cm}
    \caption{Sensitivity of the gradient. (The average value of five different $S'$.) }
    \label{fig:graddiff}
\end{figure}

\subsection{FitzHugh--Nagumo model}\label{subsec:fhn}
As the second example, we consider the FitzHugh--Nagumo model with $N=2$ and $\Omega=(-0.5,0.9)\times (-0.05,0.3)\subseteq \mathbb{R}^2$, which is described by Eq.~\eqref{eq:NWDS} with $F(x)=[x_1(x_1-c)(1-x_1)-x_2,\mu^{-1}(x_1-dx_2)]$ and $G_{i,k}(x,y)=[0,-0.0025(x_2-y_2)]$ for $x=[x_1,x_2],y=[y_1,y_2]\in\Omega$.
We set $\mu=30$, $c=-0.1$, and $d=0.5$.
We set the parameters $T$, $T_1$, and $T_2$ as the same values as in Subsection~\ref{subsec:vdp} and set $\Delta t=0.01$.
We generated the time-series $S=\{x_0,\ldots,x_T\}$.
We first compared the sensitivity $\Vert \nabla L(S)-\nabla L(S')\Vert/\Vert \nabla L(S)\Vert$ of the gradient for the loss functions defined in Subsections~\ref{subsec:practical} (KGME), \ref{subsec:pow} (called ``Power''), and \ref{subsec:fou} (called ``Fourier'').
The result for $M=3$ is shown in Table~\ref{tab:graddiff_fhn}.
We can see that the KGME approach achieves the smallest sensitivity of the gradient compared to the Power and Fourier approaches.

\begin{figure}
    \centering
    \subfigure[KGME]{\includegraphics[trim=0 30 0 0, width=0.45\linewidth]{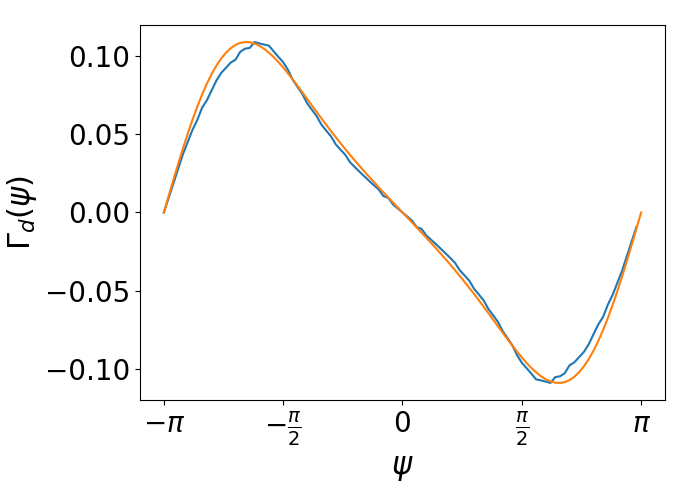}}
    \subfigure[Power]{\includegraphics[trim=0 30 0 0,width=0.45\linewidth]{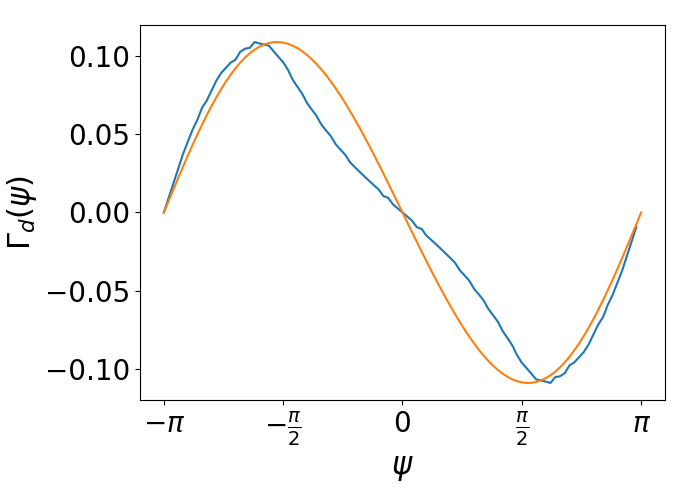}}\vspace{-.3cm}
    \subfigure[Fourier]{\includegraphics[trim=0 30 0 0,width=0.45\linewidth]{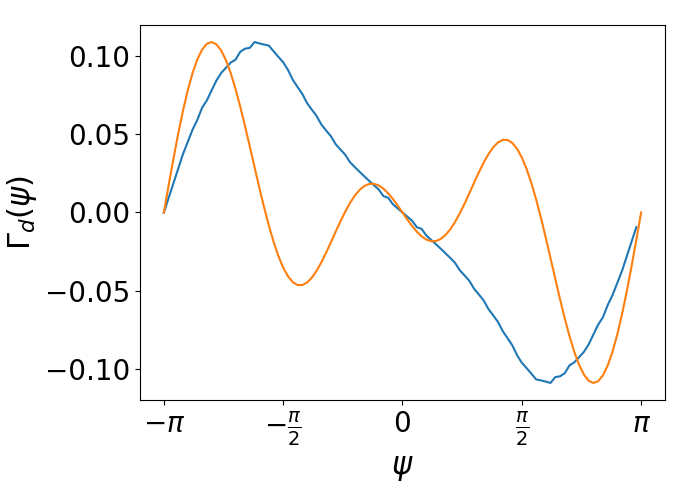}}\vspace{-.2cm}
    \caption{The function $\Gamma_d$ computed from the estimated phase coupling function. The blue line is computed by the exact model using the Fourier averaging approach~\cite{mauroy12}}
    \label{fig:phase_coupling}
\end{figure}

Next, we added observation noise to the generated time-series as $\tilde{x}_t=x_t+\xi_t$, where $\xi_t$ is randomly and independently drawn from the normal distribution with mean 0 and standard deviation 0.0001.
Then, we estimated the phase coupling function $\Gamma_i$ using the noisy data $\tilde{S}=\{\tilde{x}_0,\ldots,\tilde{x}_T\}$ with the proposed KGME approach with $M=3$.
To analyze the time evolution of the phase difference between two oscillators, the difference between the phase coupling functions is considered (See, for example, \citet{nakao16}).
Based on Eq.~\eqref{eq:phase_coupling_discretized}, the phase difference $\psi=\theta_{1}-\theta_{2}$ obeys
\begin{equation*}
    \psi(t+\Delta t)= \psi(t)+\Delta t\Gamma_d(\psi(t)),
\end{equation*}
where $\Gamma_d(\psi)={\Gamma}_1(-\psi)-{\Gamma}_2(\psi)$.
The function $\Gamma_d$ describes the synchronization; if the sign of $\psi$ and $\Gamma_d(\psi)$ is different, $\Gamma_d(\psi)$ has an effect on letting the phase difference $\psi$ small.
If the magnitude of $\Gamma_d(\psi)$ is large at $\psi$, then the effect of the interactions on synchronization is large when the phase difference is $\psi$.
We computed $\Gamma_d$ using the estimated phase coupling function.
For comparison, we also used the approaches mentioned in Subsections~\ref{subsec:pow} and \ref{subsec:fou} with $M=3$.
For optimization, we used the gradient descent method with a learning rate of 0.1.
The results after 3000 iterations are illustrated in Figure~\ref{fig:phase_coupling}.
We can see that with the KGME approach, we can estimate a phase coupling function similar to the one calculated from the exact model, compared to other approaches.

\begin{table}[t]
    \centering
    \caption{Sensitivity the gradient. (The average value $\pm$ standard deviation of five different $S'$.)}
    \label{tab:graddiff_fhn}
    \begin{tabular}{c|c}
     \hline
     Approach    &  Sensitivity \\
     \hline
     {\bf KGME (Ours)} & {\bf 0.00127$\pm$5.28$\times$1e-5}\\
     Power &  0.00148$\pm$ 4.89$\times$1e-5\\
     Fourier & 0.0955$\pm$0.0761\\
    \hline
    \end{tabular}
\end{table}

\subsection{Real-world data of cohabiting animals}\label{subsec:mice}
\begin{figure}
    \centering
    \subfigure[Time-series for mice SC8.2 and SC14.4 (1 and 3)]{\includegraphics[trim=0 30 0 0, width=0.45\linewidth]{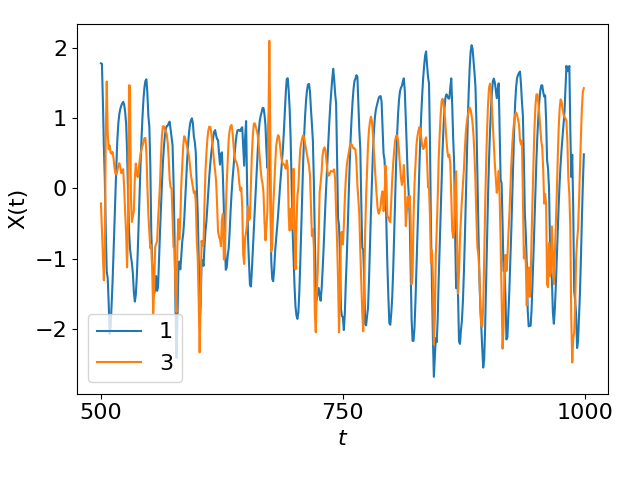}\vspace{-.5cm}}\quad
    \subfigure[Time-series for mice SC8.2 and SC21.5 (1 and 4)]{\includegraphics[trim=0 30 0 0, width=0.45\linewidth]{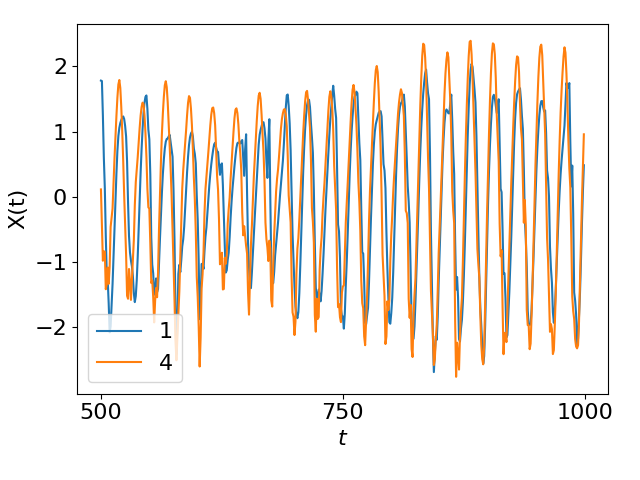}}
    \subfigure[$\Gamma_d^{1,3}$ for mice SC8.2 and SC14.4 (1 and 3)]{\includegraphics[trim=0 30 0 0, width=0.45\linewidth]{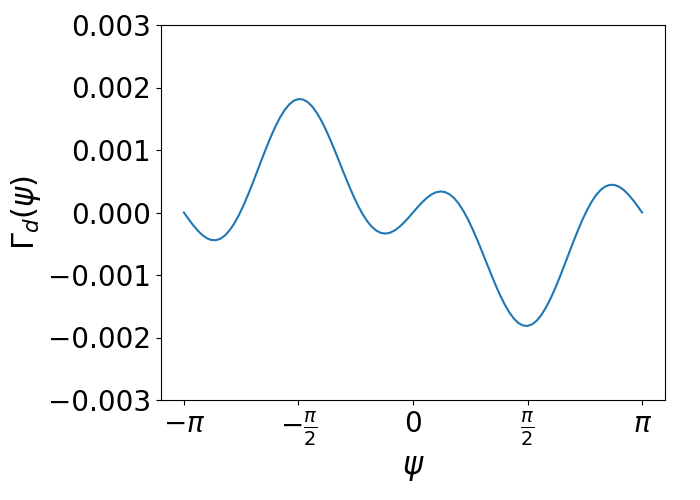}}\quad
    \subfigure[$\Gamma_d^{1,4}$ for mice SC8.2 and SC21.5 (1 and 4)]{\includegraphics[trim=0 30 0 0, width=0.45\linewidth]{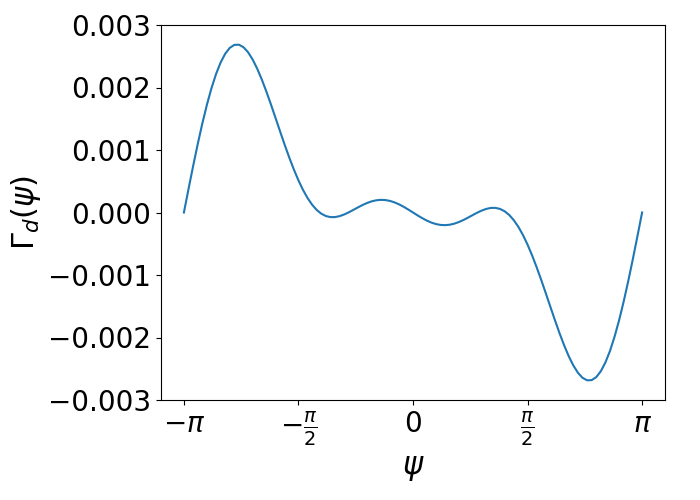}}\vspace{-.2cm}
    \caption{The time-series data and the function $\Gamma_d$ computed from the estimated phase coupling function.}
    \label{fig:mice}
\end{figure}

As the last example, we consider real-world time-series data describing the body temperature of cohabiting mice~\cite{paul15}.
Five mice were initially housed in a 12 hours light and 12 hours dark (LD) cycle.
Two mice (IDs SC8.2 and SC11.3) were housed in one of two opposite LD cycles (L:D), and the other three mice (IDs SC14.4, SC21.5, and SC25.6) were housed in the other LD cycle (D:L).
Then, they began cohabiting.
According to the results of \citet[Figure 1 (b)]{paul15}, the cohabitation caused the synchronization of their circadian rhythms.
The body temperatures of five mice were recorded every 15 minutes for 85 days (approximately 2045 hours).
We averaged every four consecutive observations to obtain the averaged body temperatures in every 1 hour.
Then, in the same manner as \citet{paul15}, we transformed the raw data using the complex-valued Morlet wavelet function.
The transformed time-series data for the pairs (SC8.2, SC14.4) and (SC8.2, SC21.5) is shown in Figure~\ref{fig:mice} (a), (b).
For $i=1,\ldots,5$, we set $X_i(t)\in\mathbb{R}^2$ as a vector describing the body temperature and its time derivative of mouse $i$ (See Table~\ref{tab:id_index}).
Thus, $N=5$ and $\mcl{X}=\mathbb{R}^2$ in this example.
We applied the proposed KGME approach with $T=2045$, $T_1=1000$, $T_2=2045$, and $M=3$.
We obtained the phase coupling function $\Gamma_i$ for $i=1,\ldots,5$.
Then, for each pair $(i,k)$, we computed the difference between $\Gamma_i$ and $\Gamma_k$ by setting $a^j_{l,m}=0$ for $(l,m)\neq (i,k)$ and focusing only on the effect between $i$ and $k$ as $\Gamma_d^{i,k}(\psi):=\tilde{\Gamma}_{i,k}(-\psi)-\tilde{\Gamma}_{k,i}(\psi)$, where $\tilde{\Gamma}_{i,k}(\psi)=2/(M(M+1))\sum_{j=1}^M\opn{arg}(a_{i,i}^j+a_{i,k}^j\mr{e}^{-\sqrt{-1}j\psi})$.
The results are illustrated in Figure~\ref{fig:mice} (c), (d).
For the pair (SC8.2, SC14.4), the oscillations of their temperatures do not completely overlap, but for the pair (SC8.2, SC21.5), the oscillations completely overlap after a sufficiently long time.
We can see that for the pair (SC8.2, SC21.5), the magnitude of $\Gamma_d^{1,4}$ is larger than $\Gamma_d^{1,3}$ for the pair (SC8.2, SC14.4).
In addition, the maximal value of $\Gamma^{1,4}(\psi)$ is obtained with a large phase difference $-\psi$, which implies that in the early stage, the effect of the interactions is greater for the pair (SC8.2, SC21.5) than for the pair (SC8.2, SC14.4). 
The result indicates that the KGME approach appropriately captures the interactions between cohabiting mice.

\begin{table}[t]
\caption{The correspondence of the index and the animal ID}\label{tab:id_index}
\centering
\begin{tabular}{ccc}
    \hline
     Index & Animal ID & LD cycle\\
     \hline
     1& SC8.2& L:D\\
     2& SC11.3& L:D\\
     3& SC14.4& D:L\\
     4& SC21.5& D:L\\
     5& SC25.6& D:L\\
     \hline
\end{tabular}
\end{table}

\section{Conclusion}
We proposed the KGME approach for estimating the phase model of coupled nonlinear oscillators.
The phase model has an important role for analyzing synchronization, which has been widely investigated in nonlinear science.
We showed that using multiple eigenvalues and eigenvectors of the Koopman operator, the proposed KGME approach is more stable with respect to the input data perturbation compared to naive and existing approaches.
The KGME approach opens up a new door for the connection between operator-theoretic data analysis and nonlinear science.

\section*{Acknowledgements}
We would like to express deep appreciation to Dr. Isao Ishikawa for constructive discussions. This work was partially supported by JST CREST Grant Number JPMJCR1913.

\bibliography{koopmanbib}

\begin{thebibliography}{48}
\providecommand{\natexlab}[1]{#1}
\providecommand{\url}[1]{\texttt{#1}}
\expandafter\ifx\csname urlstyle\endcsname\relax
  \providecommand{\doi}[1]{doi: #1}\else
  \providecommand{\doi}{doi: \begingroup \urlstyle{rm}\Url}\fi

\bibitem[Acebr\'on et~al.(2005)Acebr\'on, Bonilla, P\'erez~Vicente, Ritort, and
  Spigler]{acebron05}
Acebr\'on, J.~A., Bonilla, L.~L., P\'erez~Vicente, C.~J., Ritort, F., and
  Spigler, R.
\newblock The kuramoto model: A simple paradigm for synchronization phenomena.
\newblock \emph{Reviews of Modern Physics}, 77:\penalty0 137--185, 2005.

\bibitem[Atkinson(1972)]{atkinson72}
Atkinson, F.~V.
\newblock \emph{Multiparameter Eigenvalue Problems}.
\newblock Mathematics in Science and Engineering v. 82. Academic Press, New
  York, 1972.

\bibitem[Atkinson \& Mingarelli(2011)Atkinson and Mingarelli]{atkinson11}
Atkinson, F.~V. and Mingarelli, A.~B.
\newblock \emph{Multiparameter Eigenvalue Problems: {S}turm-{L}iouville
  Theory}.
\newblock CRC Press, Boca Raton, Florida, 2011.

\bibitem[Azencot et~al.(2020)Azencot, Erichson, Lin, and Mahoney]{azencot20}
Azencot, O., Erichson, N.~B., Lin, V., and Mahoney, M.
\newblock Forecasting sequential data using consistent {K}oopman autoencoders.
\newblock In \emph{Proceedings of the 37th International Conference on Machine
  Learning (ICML)}, 2020.

\bibitem[Bamieh(2022)]{bamieh22}
Bamieh, B.
\newblock A tutorial on matrix perturbation theory (using compact matrix
  notation).
\newblock arXiv:2002.05001v2, 2022.

\bibitem[Blaha et~al.(2011)Blaha, Pikovsky, Rosenblum, Clark, Rusin, and
  Hudson]{Blaha11}
Blaha, K.~A., Pikovsky, A., Rosenblum, M., Clark, M.~T., Rusin, C.~G., and
  Hudson, J.~L.
\newblock Reconstruction of two-dimensional phase dynamics from experiments on
  coupled oscillators.
\newblock \emph{Physics Revew E}, 84:\penalty0 046201, 2011.

\bibitem[Brown et~al.(2004)Brown, Moehlis, and Holmes]{brown04}
Brown, E., Moehlis, J., and Holmes, P.
\newblock On the phase reduction and response dynamics of neural oscillator
  populations.
\newblock \emph{Neural Computation}, 16\penalty0 (4):\penalty0 673--715, 2004.

\bibitem[Budi\v{s}i\'{c} et~al.(2012)Budi\v{s}i\'{c}, Mohr, and
  Mezi\'{c}]{mezic12}
Budi\v{s}i\'{c}, M., Mohr, R., and Mezi\'{c}, I.
\newblock Applied {K}oopmanism.
\newblock \emph{Chaos}, 22:\penalty0 047510, 2012.

\bibitem[Chen et~al.(2018)Chen, Rubanova, Bettencourt, and Duvenaud]{chen18}
Chen, R. T.~Q., Rubanova, Y., Bettencourt, J., and Duvenaud, D.~K.
\newblock Neural ordinary differential equations.
\newblock In \emph{Proceedings of the 32rd Conference on Neural Information
  Processing Systems (NeurIPS)}, 2018.

\bibitem[{\v{C}rnjari\'{c}-\v{Z}ic} et~al.(2020){\v{C}rnjari\'{c}-\v{Z}ic},
  Ma\'{c}e\v{s}i\'{c}, and Mezi\'{c}]{mezic17}
{\v{C}rnjari\'{c}-\v{Z}ic}, N., Ma\'{c}e\v{s}i\'{c}, S., and Mezi\'{c}, I.
\newblock Koopman operator spectrum for random dynamical systems.
\newblock \emph{Journal of Nonlinear Science}, 30:\penalty0 2007--2056, 2020.

\bibitem[Fujii \& Kawahara(2019)Fujii and Kawahara]{fujii19}
Fujii, K. and Kawahara, Y.
\newblock Dynamic mode decomposition in vector-valued reproducing kernel
  {H}ilbert spaces for extracting dynamical structure among observables.
\newblock \emph{Neural Networks}, 117:\penalty0 94--103, 2019.

\bibitem[Hashimoto et~al.(2020)Hashimoto, Ishikawa, Ikeda, Matsuo, and
  Kawahara]{hashimoto19}
Hashimoto, Y., Ishikawa, I., Ikeda, M., Matsuo, Y., and Kawahara, Y.
\newblock Krylov subspace method for nonlinear dynamical systems with random
  noise.
\newblock \emph{Journal of Machine Learning Research}, 21:\penalty0 172, 2020.

\bibitem[Hashimoto et~al.(2021)Hashimoto, Ishikawa, Ikeda, Komura, Katsura, and
  Kawahara]{hashimoto20}
Hashimoto, Y., Ishikawa, I., Ikeda, M., Komura, F., Katsura, T., and Kawahara,
  Y.
\newblock Reproducing kernel hilbert {$C^*$}-module and kernel mean embeddings.
\newblock \emph{Journal of Machine Learning Research}, 22:\penalty0 276, 2021.

\bibitem[Hu \& Lan(2020)Hu and Lan]{hu20}
Hu, J. and Lan, Y.
\newblock {K}oopman analysis in oscillator synchronization.
\newblock \emph{Physical Review E}, 102:\penalty0 062216, 2020.

\bibitem[Ishikawa et~al.(2018)Ishikawa, Fujii, Ikeda, Hashimoto, and
  Kawahara]{ishikawa18}
Ishikawa, I., Fujii, K., Ikeda, M., Hashimoto, Y., and Kawahara, Y.
\newblock Metric on nonlinear dynamical systems with {P}erron-{F}robenius
  operators.
\newblock In \emph{Proceedings of the 32nd Conference on Neural Information
  Processing Systems (NeurIPS)}, 2018.

\bibitem[Kadri et~al.(2016)Kadri, Duflos, Preux, Canu, Rakotomamonjy, and
  Audiffren]{kadri16}
Kadri, H., Duflos, E., Preux, P., Canu, S., Rakotomamonjy, A., and Audiffren,
  J.
\newblock Operator-valued kernels for learning from functional response data.
\newblock \emph{Journal of Machine Learning Research}, 17\penalty0
  (20):\penalty0 1--54, 2016.

\bibitem[Kawahara(2016)]{kawahara16}
Kawahara, Y.
\newblock Dynamic mode decomposition with reproducing kernels for {K}oopman
  spectral analysis.
\newblock In \emph{Proceedings of the 30th Conference on Neural Information
  Processing Systems (NIPS)}, 2016.

\bibitem[Klus et~al.(2020{\natexlab{a}})Klus, N\"{u}ske, and
  Hamzi]{klus20_generator}
Klus, S., N\"{u}ske, F., and Hamzi, B.
\newblock Kernel-based approximation of the {K}oopman generator and
  {S}chr\"{o}dinger operator.
\newblock \emph{Entropy}, 22\penalty0 (7), 2020{\natexlab{a}}.

\bibitem[Klus et~al.(2020{\natexlab{b}})Klus, N\"{u}ske, Peitz, Niemann,
  Clementi, and Sch\"{u}tte]{klus20}
Klus, S., N\"{u}ske, F., Peitz, S., Niemann, J.-H., Clementi, C., and
  Sch\"{u}tte, C.
\newblock Data-driven approximation of the {K}oopman generator: Model
  reduction, system identification, and control.
\newblock \emph{Physica D}, 406:\penalty0 132416, 2020{\natexlab{b}}.

\bibitem[Koopman(1931)]{koopman31}
Koopman, B.~O.
\newblock Hamiltonian systems and transformation in {H}ilbert space.
\newblock \emph{Proceedings of the National Academy of Sciences}, 17\penalty0
  (5):\penalty0 315--318, 1931.

\bibitem[Kovachki et~al.(2023)Kovachki, Li, Liu, Azizzadenesheli, Bhattacharya,
  Stuart, and Anandkumar]{kovachki23}
Kovachki, N., Li, Z., Liu, B., Azizzadenesheli, K., Bhattacharya, K., Stuart,
  A., and Anandkumar, A.
\newblock Neural operator: Learning maps between function spaces with
  applications to {PDE}s.
\newblock \emph{Journal of Machine Learning Research}, 24\penalty0
  (89):\penalty0 1--97, 2023.

\bibitem[Kralemann et~al.(2013)Kralemann, Fr\"{u}hwirth, Pikovsky, Rosenblum,
  Kenner, Schaefer, and Moser]{kralemann13}
Kralemann, B., Fr\"{u}hwirth, M., Pikovsky, A., Rosenblum, M., Kenner, T.,
  Schaefer, J., and Moser, M.
\newblock In vivo cardiac phase response curve elucidates human respiratory
  heart rate variability.
\newblock \emph{Nature Communications}, 4:\penalty0 2418, 2013.

\bibitem[Kuramoto(1984)]{kuramoto84}
Kuramoto, Y.
\newblock \emph{Chemical Oscillations, Waves, and Turbulence}.
\newblock Springer, New York, 1984.

\bibitem[Kutz(2013)]{kutz13}
Kutz, J.~N.
\newblock \emph{Data-Driven Modeling \& Scientific Computation: Methods for
  Complex Systems \& Big Data}.
\newblock OUP Oxford, 2013.

\bibitem[Li et~al.(2021)Li, Kovachki, Azizzadenesheli, liu, Bhattacharya,
  Stuart, and Anandkumar]{li21fourier}
Li, Z., Kovachki, N.~B., Azizzadenesheli, K., liu, B., Bhattacharya, K.,
  Stuart, A., and Anandkumar, A.
\newblock Fourier neural operator for parametric partial differential
  equations.
\newblock In \emph{Proceedings of the 9th International Conference on Learning
  Representations (ICLR)}, 2021.

\bibitem[Liu et~al.(2023)Liu, Li, Wang, and Long]{liu23}
Liu, Y., Li, C., Wang, J., and Long, M.
\newblock Koopa: Learning non-stationary time series dynamics with {K}oopman
  predictors.
\newblock In \emph{Proceedings of the 37th Conference on Neural Information
  Processing Systems (NeurIPS)}, 2023.

\bibitem[Mauroy \& Mezi\'{c}(2012)Mauroy and Mezi\'{c}]{mauroy12}
Mauroy, A. and Mezi\'{c}, I.
\newblock On the use of fourier averages to compute the global isochrons of
  (quasi) periodic dynamics.
\newblock \emph{Chaos}, 22\penalty0 (3):\penalty0 033112, 2012.

\bibitem[Mauroy \& Mezi\'{c}(2018)Mauroy and Mezi\'{c}]{mauroy18}
Mauroy, A. and Mezi\'{c}, I.
\newblock Global computation of phase-amplitude reduction for limit-cycle
  dynamics.
\newblock \emph{Chaos}, 28\penalty0 (7):\penalty0 073108, 2018.

\bibitem[Mauroy et~al.(2013)Mauroy, Mezić, and Moehlis]{mauroy13}
Mauroy, A., Mezić, I., and Moehlis, J.
\newblock Isostables, isochrons, and {K}oopman spectrum for the action–angle
  representation of stable fixed point dynamics.
\newblock \emph{Physica D}, 261:\penalty0 19--30, 2013.

\bibitem[Mezi\'{c}(2005)]{mezic05}
Mezi\'{c}, I.
\newblock Spectral properties of dynamical systems, model reduction and
  decompositions.
\newblock \emph{Nonlinear Dynamics}, 41:\penalty0 309--325, 2005.

\bibitem[Minh et~al.(2016)Minh, Bazzani, and Murino]{quang16}
Minh, H.~Q., Bazzani, L., and Murino, V.
\newblock A unifying framework in vector-valued reproducing kernel {H}ilbert
  spaces for manifold regularization and co-regularized multi-view learning.
\newblock \emph{Journal of Machine Learning Research}, 17\penalty0
  (25):\penalty0 1--72, 2016.

\bibitem[Nakao(2016)]{nakao16}
Nakao, H.
\newblock Phase reduction approach to synchronisation of nonlinear oscillators.
\newblock \emph{Contemporary Physics}, 57\penalty0 (2):\penalty0 188--214,
  2016.

\bibitem[Nakao et~al.(2014)Nakao, Yanagita, and Kawamura]{nakao14}
Nakao, H., Yanagita, T., and Kawamura, Y.
\newblock Phase-reduction approach to synchronization of spatiotemporal rhythms
  in reaction-diffusion systems.
\newblock \emph{Physical Review X}, 4:\penalty0 021032, 2014.

\bibitem[Paul et~al.(2015)Paul, Indic, and Schwartz]{paul15}
Paul, M.~J., Indic, P., and Schwartz, W.~J.
\newblock Social synchronization of circadian rhythmicity in female mice
  depends on the number of cohabiting animals.
\newblock \emph{Biology Letters}, 11\penalty0 (6):\penalty0 20150204, 2015.

\bibitem[Pietras \& Daffertshofer(2019)Pietras and Daffertshofer]{Pietras19}
Pietras, B. and Daffertshofer, A.
\newblock Network dynamics of coupled oscillators and phase reduction
  techniques.
\newblock \emph{Physics Reports}, 819:\penalty0 1--105, 2019.

\bibitem[Pikovsky et~al.(2001)Pikovsky, Rosenblum, and Kurths]{pikovsky01}
Pikovsky, A., Rosenblum, M., and Kurths, J.
\newblock \emph{Synchronization -- A Universal Concept in Nonlinear Sciences}.
\newblock Cambridge University Press, 2001.

\bibitem[Rowley et~al.(2009)Rowley, Mezi\'{c}, Bagheri, Schlatter, and
  Henningson]{rowley09}
Rowley, C.~W., Mezi\'{c}, I., Bagheri, S., Schlatter, P., and Henningson, D.~S.
\newblock Spectral analysis of nonlinear flows.
\newblock \emph{Journal of Fluid Mechanics}, 641:\penalty0 115--127, 2009.

\bibitem[Sajadi et~al.(2022)Sajadi, Kenyon, and Hodge]{sajadi22}
Sajadi, A., Kenyon, R.~W., and Hodge, B.-M.
\newblock Synchronization in electric power networks with inherent
  heterogeneity up to 100\% inverter-based renewable generation.
\newblock \emph{Nature Communications}, 13:\penalty0 2490, 2022.

\bibitem[Sch\"{o}lkopf \& Smola(2001)Sch\"{o}lkopf and Smola]{scholkopf01}
Sch\"{o}lkopf, B. and Smola, A.~J.
\newblock \emph{Learning with Kernels: Support Vector Machines, Regularization,
  Optimization, and Beyond}.
\newblock MIT Press, Cambridge, MA, USA, 2001.

\bibitem[Shirasaka et~al.(2017)Shirasaka, Kurebayashi, and Nakao]{shirasaka17}
Shirasaka, S., Kurebayashi, W., and Nakao, H.
\newblock Phase-amplitude reduction of transient dynamics far from attractors
  for limit-cycling systems.
\newblock \emph{Chaos}, 27\penalty0 (2):\penalty0 023119, 2017.

\bibitem[Stankovski et~al.(2015)Stankovski, Ticcinelli, McClintock, and
  Stefanovska]{stankovski15}
Stankovski, T., Ticcinelli, V., McClintock, P. V.~E., and Stefanovska, A.
\newblock Coupling functions in networks of oscillators.
\newblock \emph{New Journal of Physics}, 17:\penalty0 035002, 2015.

\bibitem[Takeishi et~al.(2017)Takeishi, Kawahara, and Yairi]{takeishi17}
Takeishi, N., Kawahara, Y., and Yairi, T.
\newblock Subspace dynamic mode decomposition for stochastic {K}oopman
  analysis.
\newblock \emph{Physical Review E}, 96:\penalty0 033310, 2017.

\bibitem[Teshima et~al.(2020)Teshima, Ishikawa, Tojo, Oono, Ikeda, and
  Sugiyama]{teshima20}
Teshima, T., Ishikawa, I., Tojo, K., Oono, K., Ikeda, M., and Sugiyama, M.
\newblock Coupling-based invertible neural networks are universal
  diffeomorphism approximators.
\newblock In \emph{Proceedings of the 34th Conference on Neural Information
  Processing Systems (NeurIPS)}, 2020.

\bibitem[Tokuda et~al.(2019)Tokuda, Levnajic, and Ishimura]{tokuda19}
Tokuda, I., Levnajic, Z., and Ishimura, K.
\newblock A practical method for estimating coupling functions in complex
  dynamical systems.
\newblock \emph{Philosophical Transactions of the Royal Society A:
  Mathematical, Physical and Engineering Sciences}, 377\penalty0
  (2160):\penalty0 20190015, 2019.

\bibitem[Wang et~al.(2023)Wang, Dong, Arik, and Yu]{wang23_koopman}
Wang, R., Dong, Y., Arik, S.~O., and Yu, R.
\newblock {K}oopman neural operator forecaster for time-series with temporal
  distributional shifts.
\newblock In \emph{Proceedings of the 11th International Conference on Learning
  Representations (ICLR)}, 2023.

\bibitem[Watanabe et~al.(2019)Watanabe, Kato, Shirasaka, and Nakao]{watanabe19}
Watanabe, N., Kato, Y., Shirasaka, S., and Nakao, H.
\newblock Optimization of linear and nonlinear interaction schemes for stable
  synchronization of weakly coupled limit-cycle oscillators.
\newblock \emph{Physical Review E}, 100:\penalty0 042205, 2019.

\bibitem[Williams et~al.(2015)Williams, Kevrekidis, and Rowley]{williams15}
Williams, M., Kevrekidis, I., and Rowley, C.
\newblock A data–driven approximation of the {K}oopman operator: extending
  dynamic mode decomposition.
\newblock \emph{Journal of Nonlinear Science}, 25:\penalty0 1307--1346, 2015.

\bibitem[Winfree(2001)]{winfree01}
Winfree, A.~T.
\newblock \emph{The Geometry of Biological Time}.
\newblock Springer, 2nd edition, 2001.

\end{thebibliography}
\bibliographystyle{icml2025}

\newpage
\appendix
\onecolumn
\allowdisplaybreaks
\section*{Appendix}

\section{Estimation of Koopman operators}\label{ap:Koopman_estimation}
An advantage of using Koopman operators is that we can estimate them only with given data.
In this paper, we focus on Koopman operators on a vector-valued reproducing kernel space (vvRKHS)~\cite{kadri16,quang16}.
\subsection{Vector-valued Reproducing Kernel Hilbert Space}\label{ap:vvrkhs}
We first review the theory of vvRKHSs.
To construct a vvRKHS on $\Omega^N$, we begin by a positive definite kernel.
A map $k:\Omega^N\times \Omega^N\to\mat$ is called a $\mat$-valued {\em positive definite kernel} if it satisfies the following conditions:
\begin{enumerate}
    \item $k(x,y)=k(y,x)^*$\; for $x,y\in\Omega^N$,
    \item $\sum_{i,j=1}^{n}d_i^*k(x_i,x_j)d_j\ge 0$\; for $n\in\mathbb{N}$, $d_i\in\mathbb{C}^N$, $x_i\in\Omega^N$.
\end{enumerate}
Here, $^*$ represents the adjoint.
For $x\in\Omega^N$, let $\phi_{x}$ be a $\mat$-valued map on $\Omega^N$ defined as $\phi_{x}=k(\cdot,x)$.

We construct the following vector-valued function space: 
\begin{equation*}
\hil_{0}:=\bigg\{\sum_{i=1}^n\phi_{x_i}d_i\bigg|\ n\in\mathbb{N},\ d_i\in\mathbb{C}^N,\ x_i\in\Omega^N\bigg\}.
\end{equation*}
Then, we define a map $\blacket{\cdot,\cdot}:\hil_{0}\times \hil_{0}\to\mathbb{C}$ as
\begin{align*}
&\blacketg{\sum_{i=1}^n\phi_{x_i}d_i,\sum_{j=1}^l\phi_{{y}_j}{h}_j}=\sum_{i=1}^n\sum_{j=1}^ld_i^*k(x_i,{y}_j){h}_j.
\end{align*}
By the above two properties of $\Phi$, the map $\blacket{\cdot,\cdot}$ is well-defined, satisfies the axiom of inner products, and has the reproducing property, that is,
\begin{equation*}
\blacket{\phi_{x}d,v}=d^*v(x),
\end{equation*}
for $v\in\hil_{0}$, $x\in\Omega^N$, and $d\in\mathbb{C}^N$.
The completion of $\hil_0$ 
is called the {\em vector-valued reproducing kernel Hilbert space (vvRKHS)} associated with $k$ and denoted by $\hil$.

If $N=1$, vvRKHSs are reduced to RKHSs.
If we have an RKHS $\tilde{\hil}$ associated with a positive definite kernel $\tilde{k}:\Omega\times\Omega\to\mathbb{C}$, then we can easily construct a vvRKHS that satisfies the following condition: for any $\tilde{v}\in\tilde{\hil}$, the map $v:\Omega^N\to\mathbb{C}^N$ defined as $v(x_1,\ldots,x_N)=[\tilde{v}(x_1),\ldots,\tilde{v}(x_N)]$ is in $\hil$.
Indeed, let $k:\Omega^N\times\Omega^N\to\mat$ be the $\mat$-valued positive definite kernel defined as $k(x,y)_{i,j}=\tilde{k}(x_i,y_j)$ for $x=[x_1,\ldots,x_N],y=[y_1,\ldots,y_N]\in\Omega^N$.
Let $\hil$ be the vvRKHS associated with $k$.
For any $\tilde{v}\in\tilde{\hil}_0$, $\tilde{v}$ is written as $\tilde{v}=\sum_{i=1}^n\tilde{\phi}_{z_i}c_i$ for some $n\in\mathbb{N}$, $z_i\in\Omega$, and $c_i\in\mathbb{C}$.
Then, we have
\begin{align*}
v(x)=\bigg[\sum_{i=1}^n\tilde{k}(x_1,z_i)c_i,\ldots,\sum_{i=1}^n\tilde{k}(x_N,z_i)c_i\bigg]=\sum_{i=1}^nk(x,[z_i,\ldots,z_i])c_i\mathbf{1}/N,
\end{align*}
where $\mathbf{1}$ is the $N$-dimensional vector whose elements are all $1$.
Therefore, $v\in\hil_0$, and $\hil$ satisfies the above condition.

\subsection{Estimation of Koopman operators on vvRKHSs}\label{sec:Koopman_estimation}
Here, we briefly review an approach to estimate the Koopman operator.
The estimation of the Koopman operator is proposed by, e.g.~\citet{klus20_generator,hashimoto20}.
Let $x_0,\ldots,x_T$ be given data in $\Omega^N$ generated by the dynamical system~\eqref{eq:NWDS} at time $t_0,\ldots,t_T$ with a constant time interval $\Delta t$, i.e., $t_{i+1}=t_i+\Delta t$. 
We assume that $L$ sequences $\{x^1_0,\ldots,x^1_T\},\ldots,\{x^L_0,\ldots,x^L_T\}$ of data are given.
We denote the set of these sequences by $S$.
%

We generate a subspace of $\hil$ from the given data and estimate the Koopman operator $K$ on the subspace.
For $s=0,\ldots,T$ and $i=1,\ldots,N$, let $\eta_{s,i}=1/L\sum_{j=1}^L\phi_{x^j_s}{e}_i$,
where ${e}_i$ is the vector whose $i$th element is $1$ and all the other elements are $0$, and $\phi_{x}=k(\cdot,x)$ is defined in Subsection~\ref{ap:vvrkhs}.
Let $V_T$ be the subspace of $\hil$ spanned by $\{\eta_{s,i}\,\mid\,s=0,\ldots,T-1,\ i=1,\ldots,N\}$.
Moreover, to reduce noise and extract crucial information from observed data, we use principal component analysis (PCA) and reduce the dimension of $V_T$.

We use the kernel PCA to obtain a subspace $V_{0,T}$ of $V_T$~\cite{scholkopf01}.
Let $\mathbf{G}\in\mathbb{C}^{NT\times NT}$ be the Gram matrix whose $(Ns+i,Nt+j)$-element is defined as $\blackets{\eta_{s,i},\eta_{t,j}}$.
Note that $\mathbf{G}$ is a Hermitian positive definite matrix.
Let $\lambda_1,\ldots,\lambda_{T'}> 0$ be the largest $T'$ eigenvalues of $\mathbf{G}$ and let $\bv_1,\ldots,\bv_{T'}\in\mathbb{C}^{NT}$ be the corresponding orthonormal eigenvectors.
Let $Q_S=H \mathbf{V}$,  
$H=[\eta_{0,1},\ldots,\eta_{0,N},\ldots, \eta_{T-1,1},\ldots,\eta_{T-1,N}]$, and $\mathbf{V}=[(\lambda_1)^{-1/2}\bv_1,\ldots,(\lambda_{T'})^{-1/2}\bv_{T'}]$.
We set $V_{0,T}$ as the space spanned by the column vectors of $Q_S$.
Here, we use the notation $Q_S$ to clarify the dependency on the given data $S$.
The estimation of the Koopman operator $K$ is obtained by projecting a vector in $\hil$ onto $V_{0,T}$, acting $K$ on the projected vector, and then projecting it back to the original space $\hil$.
The estimated operator is represented as $Q_SQ_S^*K^tQ_SQ_S^*$, and $Q_S^*K^tQ_S$ is a matrix representation of the estimated operator, which we denote by $\mathbf{K}^t_S$.

Let $\tilde{\mathbf{G}}\in\mathbb{C}^{NT\times NT}$ be the matrix whose $(Ns+i,Nt+j)$-element is defined as $\blackets{\eta_{s+1,i},\eta_{t,j}}$.
The matrix $\mathbf{K}_S=\mathbf{K}_S^{\Delta t}$ is shown to be represented by using computable matrices $\tilde{\mathbf{G}}$ and $\mathbf{V}$.
Note that as we mentioned in Subsection~\ref{subsec:koopman}, we denote by $K$ the Koopman operator $K^{\Delta t}$.
By the identities $\blackets{\eta_{s+1,i},\eta_{t,j}}=\blackets{\eta_{s,i},K\eta_{t,j}}$ and $Q_S=H\mathbf{V}$, we have
\begin{align*}
\mathbf{K}_S=\mathbf{V}^{*}H^*KH\mathbf{V}=\mathbf{V}^{*}\tilde{\mathbf{G}}\mathbf{V}.
\end{align*}

\section{Proofs}\label{ap:proofs}
We provide the proofs of the theorems, propositions, and lemmas in the main text.

\begin{mythm}[Theorem~\ref{thm:gme_phase_model}]
The solution $\{a_{i,k}^j\}$ of the problem~\eqref{eq:multiparam_eig_prod} satisfies the discretized version of the phase model~\eqref{eq:phase_model_couple},
\begin{align*}
&\theta_i(t+\Delta t)=\omega \Delta t +\theta_i(t)
+{\Delta t}\Gamma_i(\theta_i(t)-\theta_1(t),\ldots,\theta_i(t)-\theta_N(t))
\end{align*}
with $\Gamma_i$ in the form of Eq.~\eqref{eq:gamma_fourier}.
\end{mythm}
\begin{proof}
By the definitions of the Koopman operator and the operator $B_{i,k}$, we obtain
\begin{align}
&\prod_{j=1}^M u(X_i(t+\Delta t))^j
=\prod_{j=1}^M\bigg(e^{\sqrt{-1}j\omega\Delta t} \sum_{k=1}^Na_{i,k}^ju(X_k(t))^j\bigg).
\label{eq:phase_koopman}
\end{align}

Since $\vert {u}(X_i(t+\Delta t))\vert=\vert {u}(X_i(t))\vert$, we can normalize $u$ so that it satisfies $\vert {u}(X_i(t))\vert=1$ for any $t\in [0,\infty)$.
By this normalization, we have $\vert {u}(X_i(t))^j\vert=1$.
In addition, we have $j\theta_i(t)=\opn{arg}({u}(X_i(t))^j)$.
Therefore, we have
\begin{align}
    e^{\sqrt{-1}\sum_{j=1}^Mj\theta_{i}(t+\Delta t)}
    &=e^{\sqrt{-1}\sum_{j=1}^M(j\omega\Delta t+j\theta_{i}(t))}
    \prod_{j=1}^M\sum_{k=1}^N{a}_{i,k}^j e^{\sqrt{-1}j(\theta_{k}(t)-\theta_{i}(t))}.\label{eq:phase_koopman_diff} 
\end{align}

Therefore, by calculating arguments of both sides of Eq.~\eqref{eq:phase_koopman_diff}, we obtain
\begin{align*}
    &\sum_{j=1}^Mj\theta_{i}(t+\Delta t)
    \sum_{j=1}^M\!\!\bigg(\!j\omega\Delta t + j\theta_{i}(t)+ \opn{arg}\!\bigg(\!\sum_{k=1}^N{a}_{i,k}^j e^{\sqrt{-1}j(\theta_{k}(t)-\theta_{i}(t))}\!\bigg)\!\bigg).
\end{align*}
As a result, we obtain
\begin{align*}
    &\theta_{i}(t+\Delta t)
    = \omega\Delta t + \theta_{i}(t)
    \frac{2}{M(M+1)}\sum_{j=1}^M\opn{arg}\bigg(\sum_{k=1}^N{a}_{i,k}^j e^{\sqrt{-1}j(\theta_{k}(t)-\theta_{i}(t))}\bigg)
\end{align*}
Therefore, we obtain the discretized version of the phase model~\eqref{eq:phase_model_couple} by setting
\begin{align*}
&{\Gamma}_{i}(\psi_1,\ldots,\psi_N)
=\frac{1}{\Delta t} \frac{2}{M(M+1)}\sum_{j=1}^M\opn{arg}\bigg(\sum_{k=1}^N{a}_{i,k}^j e^{-\sqrt{-1}j\psi_k}\bigg).
\end{align*}
\end{proof}

\begin{mythm}[Theorem~\ref{thm:stability}]
Assume 
$\Vert Q_{S_1}-Q_{S_1'}\Vert\le \epsilon/(2\Vert\tilde{K}\Vert)$, $\Vert Q_{S_2}-Q_{S_2'}\Vert\le \epsilon/(2\Vert {K}\Vert)$ and $\vert a^j_{S,i,k}-a^j_{S',i,k}\vert\le\epsilon$ for some $\epsilon>0$.
Assume, in addition, for any time-series $S$, $\vert A_S^j\vert\le D_A$, $\Vert U_S\Vert\le D_U$, and $\Vert W_S\Vert\le D_W$ for some $D_A,D_U,D_W>0$.
Then, we have
\begin{align*}
&\vert a^{\opn{new},j}_{S,i,k}-a^{\opn{new},j}_{S',i,k}\vert 
\le \epsilon+2\epsilon\eta N  D_U^2 (1+N+D_UD_WD_A
+2(\Vert K\Vert+D_A)D_UD_W\alpha_S)+O(\epsilon^2).
\end{align*}
\end{mythm}

\begin{proof}
Since $\Vert Q_{S_1}-Q_{S_1'}\Vert\le \epsilon/(2\Vert\tilde{K}\Vert)$, we have
\begin{align*}
\Vert \mathbf{\tilde{K}}_{S_1}-\mathbf{\tilde{K}}_{S_1'}\Vert
&=\Vert Q_{S_1}^*\tilde{K}Q_{S_1}-Q_{S_1'}^*\tilde{K}Q_{S_1'}\Vert
\le \Vert Q_{S_1}^*\tilde{K}(Q_{S_1}-Q_{S_1'})\Vert + \Vert (Q_{S_1}-Q_{S_1'})^*\tilde{K}Q_{S_1'}\Vert\\
&\le 2\Vert\tilde{K}\Vert\,\Vert Q_{S_1}-Q_{S_1'}\Vert
\le \epsilon.
\end{align*}
In the same manner, we have $\Vert \mathbf{{K}}_{S_2}-\mathbf{{K}}_{S_2'}\Vert\le \epsilon$.
By Lemma~\ref{lem:stability_eig}, for any eigenvector $\bu_{S_1}$ of $\mathbf{\tilde{K}}_{S_1}$ with $W_S^*U_S=I$ and for any $i=1,\ldots,N$, we have
\begin{align*}
\Vert \bu_{S_1}-\bu_{S_1'}\Vert
&\le \epsilon\Vert U_{S_1}\Vert^2\Vert W_{S_1}\Vert\max_{i,j}\frac{1}{\lambda_{S_1,i}-\lambda_{S_1,j}}+O(\epsilon^2),\\
\vert \lambda_{S_1,i}-\lambda_{S_1',i}\vert
&\le \epsilon\Vert U_{S_1}\Vert\Vert W_{S_1}\Vert.
\end{align*}
In the following, for simplification, we abuse the notation and denote $L^{\opn{KGME}}$ by $L$ and $L_{j}^{\opn{KGME}}$ by $L_{j}$. 
The difference between the gradients of the loss function for $S$ and $S'$ is
\begin{align}
\vert \nabla L(S)-\nabla L(S')\vert
&=\bigg\vert\sum_{j=1}^M(\nabla\Vert L_{j}(S)\Vert^2-\nabla\Vert L_{j}(S')\Vert^2)\bigg\vert\nn\\
&= \bigg\vert\sum_{j=1}^M(2{\mathbb{J} L_{j}(S)}{ L_{j}(S)}-2{\mathbb{J} L_{j}(S')} L_{j}(S'))\bigg\vert,\label{eq:loss_grad_diff}
\end{align}
where $\mathbb{J} L_{j}$ is the Jacobian of $ L_{j}$ with respect to $(a_{i,k}^j)_{i,k=1,\ldots,N}^{j=1,\ldots,M}$.
In addition, for $i,k=1,\ldots,N$ and {j=1,\ldots,M}, we have
\begin{align}
&\vert \mathbb{J} L_{j}(S) L_{j}(S)-\mathbb{J} L_{j}(S') L_{j}(S')\vert_{i,j,k}\le
\vert \mathbb{J}{L}_{j}(S)({L}_{j}(S)-{L}_{j}(S'))\vert_{i,j,k} + \vert (\mathbb{J}{L}_{j}(S)-\mathbb{J}{L}_{j}(S')) L_{j}(S')\vert_{i,j,k}\nn\\
&\le
\Vert \mathbb{J}{L}_{j}(S)\Vert \Vert{L}_{j}(S)-{L}_{j}(S')\Vert + \Vert \mathbb{J}{L}_{j}(S)-\mathbb{J}{L}_{j}(S')\Vert\Vert{L}_{j}(S')\Vert\nn\\
&\le \Vert [PB_{1,1}\bu_{S_1}^j,\ldots,PB_{N,N}\bu_{S_1}^j]\Vert_F\bigg(\Vert \mathbf{K}_{S_2}P\mathbf{u}_{S_1}^j-\mathbf{K}_{S_2'}P\mathbf{u}_{S_1'}^j\Vert\nn\\
&\qquad+\bigg\Vert\lambda_{S_1,j}\sum_{i,k=1}^Na^j_{S,i,k}PB_{i,k}\mathbf{u}_{S_1}^j-\lambda_{S_1',j}\sum_{i,k=1}^Na^j_{S',i,k}PB_{i,k}\mathbf{u}_{S_1'}^j\bigg\Vert\bigg)\nn\\
&\qquad +\Vert [PB_{1,1}\bu_{S_1}^j,\ldots,PB_{N,N}\bu_{S_1}^j]-[PB_{1,1}\bu_{S_1'}^j,\ldots,PB_{N,N}\bu_{S_1'}^j]\Vert_F(\Vert K\Vert\Vert \bu^j_{S_1'}\Vert+\Vert P(A_{S'}^j\otimes I)\Vert \Vert \bu^j_{S_1'}\Vert ) \nn\\
&\le N\Vert \bu_{S_1}^j\Vert \bigg(\Vert \mathbf{K}_{S_2}-\mathbf{K}_{S_2'}\Vert\Vert \bu_{S_1}^j\Vert+ \Vert \mathbf{K}_{S_2'}\Vert\Vert \mathbf{u}_{S_1}^j-\mathbf{u}_{S_1'}^j\Vert+
\bigg\Vert \sum_{i,k=1}^N\lambda_{S_1,j}a^j_{S,i,k}PB_{i,k}(\mathbf{u}_{S_1}^j-\mathbf{u}_{S_1'}^j)\bigg\Vert\nn\\
&\qquad+\bigg\Vert \sum_{i,k=1}^N(\lambda_{S_1,j}a^j_{S,i,k}-\lambda_{S_1,j}a^j_{S',i,k})PB_{i,k}\mathbf{u}_{S_1'}^j\bigg\Vert\bigg)\nn\\
&\qquad +N \Vert \bu_{S_1}^j-\bu_{S_1'}^j\Vert(\Vert K\Vert\Vert \bu^j_{S_1'}\Vert+\Vert A_{S'}^j\otimes I\Vert \Vert \bu^j_{S_1'}\Vert )\nn\\
&\le N\Vert \bu_{S_1}^j\Vert \bigg(\epsilon\Vert \bu_{S_1}^j\Vert+ \Vert K\Vert\Vert  \Vert\mathbf{u}_{S_1}^j-\mathbf{u}_{S_1'}^j\Vert+
\Vert A_{S}^j\otimes I\Vert \Vert\mathbf{u}_{S_1}^j-\mathbf{u}_{S_1'}^j\Vert
+\Vert (\lambda_{S_1,j}A_{S}^j-\lambda_{S_1',j}A^j_{S'})\otimes I\Vert\Vert \mathbf{u}_{S_1'}^j\Vert\bigg)\nn\\
&\qquad +N \Vert \bu_{S_1}^j-\bu_{S_1'}^j\Vert(\Vert K\Vert\Vert \bu^j_{S_1'}\Vert+\Vert A^j_{S'}\otimes I\Vert \Vert \bu^j_{S_1'}\Vert )\nn\\
&\le N\Vert \bu_{S_1}^j\Vert \bigg(\epsilon\Vert \bu_{S_1}^j\Vert+ (\Vert K\Vert+\Vert A_{S}^j\otimes I\Vert)  \bigg(\epsilon\Vert U_{S_1}\Vert^2\Vert W_{S_1}\Vert\max_{i,j}\frac{1}{\lambda_{S_1,i}-\lambda_{S_1,j}}+O(\epsilon^2)\bigg)\nn\\
&\qquad+N\epsilon\Vert \mathbf{u}_{S_1'}^j\Vert+(\epsilon\Vert U_{S_1}\Vert\Vert W_{S_1}\Vert+O(\epsilon^2))\Vert A_{S'}^j\otimes I\Vert \Vert \mathbf{u}_{S_1'}^j\Vert\bigg)\nn\\
&\qquad +N (\epsilon\Vert U_{S_1}\Vert^2\Vert W_{S_1}\Vert\max_{i,j}\frac{1}{\lambda_{S_1,i}-\lambda_{S_1,j}}+O(\epsilon^2))(\Vert K\Vert\Vert \bu^j_{S_1}\Vert+\Vert A_{S'}^j\otimes I\Vert \Vert \bu^j_{S_1}\Vert )\nn\\
&\le \epsilon N  D_U^2(1+N+D_UD_WD_A)
+2\epsilon N D_U(\Vert K\Vert+D_A)D_U^2D_W\max_{i,j}\frac{1}{\lambda_{S_1,i}-\lambda_{S_1,j}}+O(\epsilon^2),
\label{eq:loss_diff}
\end{align}
where $\Vert\cdot\Vert_F$ is the Frobenius norm.
Combining Eq.~\eqref{eq:loss_diff} with \eqref{eq:loss_grad_diff}, since the derivative of $L_j(S)$ with respect to $a_{i,k}^l$ for $l\neq j$ is $0$, for any $i,j,k=1,\ldots,N$, we obtain
\begin{align*}
\vert \nabla {L}(S)-\nabla {L}(S')\vert_{i,j,k}
&\le 2\epsilon N  D_U^2\bigg(1+N+D_UD_WD_A+2(\Vert K\Vert+D_A)D_UD_W\max_{i,j}\frac{1}{\lambda_{S_1,i}-\lambda_{S_1,j}}\bigg)+O(\epsilon^2).
\end{align*}
The updated parameter $a^{j,\opn{new}}_{i,k}$ is written as
\begin{align*}
a^{\opn{new},j}_{S,i,k}=a^j_{S,i,k}-\eta(\nabla L(S))_{i,k},
\end{align*}
where $\eta>0$ is the learning rate.
Thus, we have
\begin{align*}
\vert a^{\opn{new},j}_{S,i,k}-a^{\opn{new},j}_{S',i,k}\vert 
&\le\vert a^j_{S,i,k}-a^j_{S',i,k}\vert + \eta\vert \nabla L(S)-\nabla L(S')\vert_{i,k}\\
&\le \epsilon+2\epsilon\eta N  D_U^2\bigg(1+N+D_UD_WD_A+2(\Vert K\Vert+D_A)D_UD_W\max_{i,j}\frac{1}{\lambda_{S_1,i}-\lambda_{S_1,j}}\bigg)+O(\epsilon^2).
\end{align*}
\end{proof}

\begin{mythm}[Lemma~\ref{lem:stability_eig}]
Assume $\Vert \mathbf{\tilde{K}}_{S_1}-\mathbf{\tilde{K}}_{S_1'}\Vert\le \epsilon$ for some $\epsilon>0$.
Then, we have
\begin{align*}
\Vert U_{S_1}-U_{S_1'}\Vert 
&\le \epsilon\Vert U_{S_1}\Vert^2\Vert W_{S_1}\Vert\alpha_{S}+O(\epsilon^2)\\
\vert\lambda_{S_1,i}-\lambda_{S_1',i}\vert&\le \epsilon\Vert U_{S_1}\Vert\Vert W_{S_1}\Vert+O(\epsilon^2).
\end{align*}
\end{mythm}

\begin{proof}
By Section 3.1 in~\citet{bamieh22}, we have
\begin{align*}
U_{S_1}-U_{S_1'}&=-\epsilon U_{S_1}\bigg(\Pi_{S_1}\odot \bigg(W_{S_1}^*\frac{1}{\epsilon}\bigg(\mathbf{\tilde{K}}_{S_1}-\mathbf{\tilde{K}}_{S_1'})U_{S_1}\bigg)\bigg)\\
&\lambda_{S_1,i}-\lambda_{S_1',i}=-\epsilon W_{S_1}^*\frac{1}{\epsilon}(\mathbf{\tilde{K}}_{S_1}-\mathbf{\tilde{K}}_{S_1'})U_{S_1},
\end{align*}
where $(\Pi_{S_1})_{i,j}=0$ if $i=j$ and $(\Pi_{S_1})_{i,j}=1/(\lambda_{S_1,i}-\lambda_{S_1,j})$ if $i\neq j$.
In addition, $\odot$ denotes the elementwise product of matrices.
Thus, we have
\begin{align*}
\Vert U_{S_1}-U_{S_1'}\Vert 
&\le \epsilon\Vert U_{S_1}\Vert^2\Vert W_{S_1}\Vert\max_{i,j}\frac{1}{\lambda_{S_1,i}-\lambda_{S_1,j}}\\
\vert \lambda_{S_1,i}-\lambda_{S_1',i}\vert&\le \epsilon\Vert U_{S_1}\Vert\Vert W_{S_1}\Vert,
\end{align*}
which completes the proof.
\end{proof}

\begin{mythm}[Proposition~\ref{prop:stability_single_eig}]
Let $\hil$ be an RKHS associated with a kernel $k$ that satisfies $\sup_{x\in\Omega}\Vert k(x,x)\Vert \le C$ for some $C>0$.
Let $\opn{Ev}_S:C(\Omega)^{N^2}\to\mathbb{C}^{N^2}$ be defined as $v\mapsto v(x)$, where $C(\Omega)$ is the Banach space of continuous functions on $\Omega$.
Assume $\Vert\mr{Ev}_{S_2}-\mr{Ev}_{S_2'}\Vert\le \epsilon$, $\Vert Q_{S_1}-Q_{S_1'}\Vert\le \epsilon/(2\Vert\tilde{K}\Vert)$, and $\vert a^j_{S,i,k}-a^j_{S',i,k}\vert\le\epsilon$ for some $\epsilon>0$.
Assume for any time-series $S$, $\Vert \opn{Ev}_S\Vert\le D_E$, $\vert A_S^j\vert\le D_A$, $\Vert U_S\Vert\le D_U$, $\Vert Q_{S_1}\bu^1(x)^j\Vert\le D_U$, and $\Vert W_S\Vert\le D_W$ for some $D_E,D_A,D_U,D_W>0$.
Then, we have
\begin{align*}
\vert a^{\opn{new},j}_{S,i,k}-a^{\opn{new},j}_{S',i,k}\vert 
&\le \epsilon+2\epsilon N  D_U^2(1+D_A+D_E +ND_E)\nn\\
&\quad +2\epsilon N  D_U^2j\bigg(D_A \Delta t
+2D_E\sqrt{N}C^{1/2}(1+D_A)\bigg(D_U^2D_W\alpha_S+\frac{D_U}{2\Vert\tilde{K}\Vert}\bigg)\bigg)
+O(\epsilon^2).
\end{align*}
\end{mythm}

\begin{proof}
By Lemma~\ref{lem:stability_eig}, for any eigenvector $\bu_{S_1}$ of $\mathbf{\tilde{K}}_{S_1}$ with $W_S^*U_S=I$ and for any $i=1,\ldots,N$, we have
\begin{align*}
\vert Q_{S_1}\bu_{S_1}(x)-Q_{S_1'}\bu_{S_1'}(x)\vert_i
&\le \Vert k(x,x)\Vert^{1/2} \Vert Q_{S_1}\bu_{S_1}-Q_{S_1'}\bu_{S_1'}\Vert\\
&\le C^{1/2}(\Vert Q_{S_1}(\bu_{S_1}-\bu_{S_1'})\Vert+\Vert(Q_{S_1}-Q_{S_1'})\bu_{S_1'}\Vert)\\
&\le C^{1/2}\bigg(\Vert U_{S_1}-U_{S_1'}\Vert +\frac{\epsilon}{\Vert 2\tilde{K}\Vert}\Vert \bu_{S_1'}\Vert\bigg)\\
&\le C^{1/2}\bigg(\epsilon\Vert U_{S_1}\Vert^2\Vert W_{S_1}\Vert\max_{i,j}\frac{1}{\lambda_{S_1,i}-\lambda_{S_1,j}}+\frac{\epsilon}{2\Vert\tilde{K}\Vert}\Vert \bu_{S_1'}\Vert\bigg)+O(\epsilon^2).
\end{align*}

Since $\vert Q_{S_1}\bu_{S_1}^1(x)\vert\le D_U$, we have
\begin{align*}
&\vert (Q_{S_1}\bu_{S_1}^1(x))^j-(Q_{S_1'}\bu_{S_1'}^1(x))^j\vert\\
&= \vert Q_{S_1}\bu_{S_1}^1(x)-Q_{S_1'}\bu_{S_1'}^1(x)\vert\,\\
&\qquad\cdot\vert (Q_{S_1}\bu_{S_1}^1(x))^{j-1}+(Q_{S_1}\bu_{S_1}^1(x))^{j-2}Q_{S_1'}\bu_{S_1'}^1(x)+\cdots +Q_{S_1}\bu_{S_1}^1(x)(Q_{S_1'}\bu_{S_1'}^1(x))^{j-2}+(Q_{S_1'}\bu_{S_1'}^1(x))^{j-1}\vert\\
&\le \vert Q_{S_1}\bu_{S_1}^1(x)-Q_{S_1'}\bu_{S_1'}^1(x)\vert D_U j.
\end{align*}

In the following, for simplification, we abuse the notation and denote $L^{\opn{pow}}$ by $L$ and $L_{j,x}^{\opn{pow}}$ by $L_{j,x}$. 
The difference between the gradients of the values of loss function for $S$ and $S'$ is
\begin{align}
\vert \nabla L(S)-\nabla L(S')\vert
&=\bigg\vert\sum_{j=1}^M\bigg(\nabla\sum_{x\in S_2} \Vert L_{j,x}(S_1)\Vert^2-\nabla\sum_{x\in S_2'}\Vert L_{j,x}(S_1')\Vert^2\bigg)\bigg\vert\nn\\
&= \bigg\vert\sum_{j=1}^M\bigg(2\sum_{x\in S_2}{\mathbb{J} L_{j,x}(S_1)}{ L_{j,x}(S_1)}-2\sum_{x\in S_2'}{\mathbb{J} L_{j,x}(S_1')} L_{j,x}(S_1')\bigg)\bigg\vert.\label{eq:loss_grad_diff_eig}
\end{align}

In addition, for $i,k=1,\ldots,N$ and $j=1,\ldots,M$, we have
\begin{align}
&\bigg\vert \sum_{x\in S_2}\mathbb{J} L_{j,x}(S_1) L_{j,x}(S_1)-\sum_{x\in S_2'}\mathbb{J} L_{j,x}(S_1') L_{j,x}(S_1')\bigg\vert_{i,j,k}
= \vert \mr{Ev}_{S_2}(\mathbb{J} L_{j,\cdot}(S_1) L_{j,\cdot}(S_1))-\mr{Ev}_{S_2'}(\mathbb{J} L_{j,\cdot}(S_1') L_{j,\cdot}(S_1'))\vert_{i,j,k}\nn\\
&\le
\Vert \mr{Ev}_{S_2}-\mr{Ev}_{S_2'}\Vert\sup_{x\in\Omega^N}\Vert \mathbb{J}{L}_{j,x}(S_1){L}_{j,x}(S_1)\Vert + \Vert \mr{Ev}_{S_2'}\Vert \sup_{x\in\Omega^N}\Vert \mathbb{J}{L}_{j,\cdot}(S_1){L}_{j,x}(S_1)-\mathbb{J}{L}_{j,x}(S_1'){L}_{j,x}(S_1')\Vert\nn\\
&\le \Vert \mr{Ev}_{S_2}-\mr{Ev}_{S_2'}\Vert\sup_{x\in\Omega^N}\Vert \mathbb{J}{L}_{j,x}(S_1){L}_{j,\cdot}(S_1)\Vert+\Vert \mr{Ev}_{S_2'}\Vert\sup_{x\in\Omega^N}\Vert \mathbb{J}{L}_{j,x}(S_1)\Vert \Vert{L}_{j,x}(S_1)-{L}_{j,x}(S_1')\Vert \nn\\
&\qquad +\Vert \mr{Ev}_{S_2'}\Vert\sup_{x\in\Omega^N}\Vert \mathbb{J}{L}_{j,x}(S_1)- \mathbb{J}{L}_{j,x}(S_1')\Vert\Vert{L}_{j,x}(S_1')\Vert\nn\\
&\le \epsilon\sup_{x\in\Omega^N}\Vert \mr{e}^{\sqrt{-1}j\omega_{S_1}\Delta t}[B_{1,1}Q_{S_1}\bu^1_{S_1}(x)^j,\ldots,B_{N,N}Q_{S_1}\bu^1_{S_1}(x)^j]\Vert_F\nn\\ &\qquad\cdot\bigg\Vert Q_{S_1}\bu_{S_1}^1(\Phi(\Delta t,x))^j-\mr{e}^{\sqrt{-1}j\omega_{S_1}\Delta t}\sum_{i,k=1}^Na^j_{S,i,k}B_{i,k}Q_{S_1}\mathbf{u}_{S_1}^1(x)^j\bigg\Vert\nn\\
&\qquad +\Vert \mr{Ev}_{S_2'}\Vert N\sup_{x\in\Omega^N}\Vert Q_{S_1}\bu^1_{S_1}(x)^j\Vert
\sup_{x\in\Omega^N}\bigg(\Vert Q_{S_1}\bu_{S_1}^1(\Phi(\Delta t,x))^j-Q_{S_1'}\bu_{S_1'}^1(\Phi(\Delta t,x))^j\Vert\nn\\
&\qquad +\bigg\Vert \mr{e}^{\sqrt{-1}j\omega_{S_1}\Delta t}\sum_{i,k=1}^Na^j_{S,i,k}B_{i,k}Q_{S_1}\mathbf{u}_{S_1}^1(x)^j-\mr{e}^{\sqrt{-1}j\omega_{S_1'}\Delta t}\sum_{i,k=1}^Na^j_{S',i,k}B_{i,k}Q_{S_1'}\mathbf{u}_{S_1'}^1(x)^j\bigg\Vert\bigg)\nn\\
&\qquad +\Vert \mr{Ev}_{S_2'}\Vert N\sup_{x\in\Omega^N}\Vert Q_{S_1}\bu^1_{S_1}(x)^j-Q_{S_1'}\bu^1_{S_1'}(x)^j\Vert(\Vert Q_{S_1'}\bu_{S_1'}^1(\Phi(\Delta t,x))^j\Vert+\Vert A_{S'}^jQ_{S_1'}\bu^1_{S_1'}(x)^j\Vert)\nn\\
&\le \epsilon N\sup_{x\in\Omega^N}\Vert Q_{S_1}\bu^1_{S_1}(x)^j\Vert (\Vert Q_{S_1}\bu_{S_1}^1(\Phi(\Delta t,x))^j\Vert+\Vert A_{S}^jQ_{S_1}\bu^1_{S_1}(x)^j\Vert)\nn\\
&\qquad +D_E N D_U\bigg(\sup_{x\in\Omega^N}\Vert Q_{S_1}\bu^1_{S_1}(x)^j-Q_{S_1'}\bu^1_{S_1'}(x)^j\Vert+\sup_{x\in\Omega^N}\bigg\Vert (\mr{e}^{\sqrt{-1}j\omega_{S_1}\Delta t}- \mr{e}^{\sqrt{-1}j\omega_{S_1'}\Delta t})\sum_{i,k=1}^Na^j_{S,i,k}B_{i,k}Q_{S_1}\mathbf{u}_{S_1}^1(x)^j\bigg\Vert\nn\\
&\qquad+\sup_{x\in\Omega^N}\bigg\Vert\mr{e}^{\sqrt{-1}j\omega_{S_1'}\Delta t}\sum_{i,k=1}^N(a^j_{S,i,k}-a^j_{S',i,k})B_{i,k}Q_{S_1}\mathbf{u}_{S_1}^1(x)^j\bigg\Vert\nn\\
&\qquad+\sup_{x\in\Omega^N}\bigg\Vert\mr{e}^{\sqrt{-1}j\omega_{S_1'}\Delta t}\sum_{i,k=1}^Na^j_{S',i,k}B_{i,k}(Q_{S_1}\mathbf{u}_{S_1}^1(x)^j-Q_{S_1'}\mathbf{u}_{S_1'}^1(x)^j)\bigg\Vert\bigg)\nn\\
&\qquad +\Vert \mr{Ev}_{S_2'}\Vert N\sup_{x\in\Omega^N}\Vert Q_{S_1}\bu^1_{S_1}(x)^j-Q_{S_1'}\bu^1_{S_1'}(x)^j\Vert(\Vert Q_{S_1'}\bu_{S_1'}^1(\Phi(\Delta t,x))^j\Vert+\Vert A_{S'}^jQ_{S_1'}\bu^1_{S_1'}(x)^j\Vert)\nn\\
&\le \epsilon ND_U (D_U+D_AD_U)+D_E N D_U(\sup_{x\in\Omega^N}\Vert Q_{S_1}\bu^1_{S_1}(x)^j-Q_{S_1'}\bu^1_{S_1'}(x)^j\Vert\nn\\
&\qquad +j\Delta t\vert\omega_{S_1}-\omega_{S_1'}\vert D_AD_U
+N\epsilon D_U+D_A\sup_{x\in\Omega^N}\Vert Q_{S_1}\bu^1_{S_1}(x)^j-Q_{S_1'}\bu^1_{S_1'}(x)^j\Vert)\nn\\
&\qquad +D_E N \sup_{x\in\Omega^N}\Vert Q_{S_1}\bu^1_{S_1}(x)^j-Q_{S_1'}\bu^1_{S_1'}(x)^j\Vert(D_U+D_AD_U)\nn\\
&\le \epsilon N D_U^2(1+D_A+D_E D_A j\Delta t+ND_E)\nn\\
&\qquad +ND_U D_E (1+D_A+1+D_A)jD_U\sqrt{N}C^{1/2}\epsilon\bigg(D_U^2D_W\max_{i,j}\frac{1}{\lambda_{S_1,i}-\lambda_{S_1',i}}+\frac{D_U}{2\Vert\tilde{K}\Vert}+O(\epsilon^2)\bigg).\label{eq:loss_diff_eig}
\end{align}

Combining Eq.~\eqref{eq:loss_diff_eig} with \eqref{eq:loss_grad_diff_eig}, for any $i,j,k=1,\ldots,N$, we obtain
\begin{align*}
\vert \nabla {L}(S)-\nabla {L}(S')\vert_{i,j,k}
&\le 2\epsilon N  D_U^2(1+D_A+D_E +ND_E)\nn\\
&\quad +2\epsilon N  D_U^2j\bigg(D_A \Delta t+2D_E\sqrt{N}C^{1/2}(1+D_A)\bigg(D_U^2D_W\max_{i,j}\frac{1}{\lambda_{S_1,i}-\lambda_{S_1,j}}+\frac{D_U}{2\Vert\tilde{K}\Vert}\bigg)\bigg)\bigg)+O(\epsilon^2).
\end{align*}
Thus, we have
\begin{align*}
\vert a^{\opn{new},j}_{S,i,k}-a^{\opn{new},j}_{S',i,k}\vert 
&\le\vert a^j_{S,i,k}-a^j_{S',i,k}\vert + \eta\vert \nabla L(S)-\nabla L(S')\vert_{i,j,k}\\
&\le \epsilon+2\epsilon \eta N  D_U^2(1+D_A+D_E +ND_E)\nn\\
&\quad +2\epsilon \eta N  D_U^2j\bigg(D_A \Delta t+2D_E\sqrt{N}C^{1/2}(1+D_A)\bigg(D_U^2D_W\max_{i,j}\frac{1}{\lambda_{S_1,i}-\lambda_{S_1,j}}+\frac{D_U}{2\Vert\tilde{K}\Vert}\bigg)\bigg)+O(\epsilon^2).
\end{align*}

\end{proof}

\begin{mythm}[Proposition~\ref{prop:stability_fou}]
Assume there exists $\epsilon>0$ such that for any $x\in\Omega$ and $i,k=1,\ldots,N$, $\vert \Theta_{S_1}(x)-\Theta_{S_1'}(x)\vert_i\le \epsilon$, $\vert\omega_{S_1}-\omega_{S_1'}\vert\le \epsilon$, and $\vert a_{S,i,k}^j-a_{S',i,k}^j\vert\le \epsilon$.
Assume for any time-series $S$, $\Vert \opn{Ev}_S\Vert\le D_E$, $\vert A_S^j\vert\le D_A$.
Then, we have
\begin{align*}
\vert a^{\opn{new},j}_{S,i,k}-a^{\opn{new},j}_{S',i,k}\vert
&\le \epsilon+2\epsilon \eta N\sqrt{M}(2\sqrt{N}+2\pi\Delta t+MD_A\sqrt{N}
+D_E(1+\Delta t+MN\sqrt{N}+2M\sqrt{N}D_Aj))\\
&\quad+2\epsilon\eta D_E N M(2\sqrt{N}+2\pi\Delta t+MD_A\sqrt{N}).
\end{align*}
\end{mythm}
\begin{proof}
In the following, for simplification, we abuse the notation and denote $L^{\opn{Fou}}$ by $L$ and $L_{1,x}^{\opn{Fou}}$ by $L_{1,x}$. 
The difference between the gradients of the values of loss function for $S$ and $S'$ is
\begin{align*}
\vert \nabla L(S)-\nabla L(S')\vert
&=\bigg\vert\nabla\sum_{x\in S_2} \Vert L_{1,x}(S_1)\Vert^2-\nabla\sum_{x\in S_2'}\Vert L_{1,x}(S_1')\Vert^2\bigg\vert\nn\\
&= \bigg\vert 2\sum_{x\in S_2}{\mathbb{J} L_{1,x}(S_1)}{ L_{1,x}(S_1)}-2\sum_{x\in S_2'}{\mathbb{J} L_{1,x}(S_1')} L_{1,x}(S_1')\bigg\vert.
\end{align*}
In addition, for $i,k=1,\ldots,N$ and $j=1,\ldots,M$, we have
\begin{align*}
&\bigg\vert \sum_{x\in S_2}\mathbb{J} L_{1,x}(S_1) L_{1,x}(S_1)-\sum_{x\in S_2'}\mathbb{J} L_{1,x}(S_1') L_{1,x}(S_1')\bigg\vert_{i,j,k}
= \vert \mr{Ev}_{S_2}(\mathbb{J} L_{1,\cdot}(S_1) L_{1,\cdot}(S_1))-\mr{Ev}_{S_2'}(\mathbb{J} L_{1,\cdot}(S_1') L_{1,\cdot}(S_1'))\vert_{i,j,k}\nn\\
&\le
\Vert \mr{Ev}_{S_2}-\mr{Ev}_{S_2'}\Vert\sup_{x\in\Omega^N}\Vert \mathbb{J}{L}_{1,x}(S_1){L}_{1,x}(S_1)\Vert + \Vert \mr{Ev}_{S_2'}\Vert \sup_{x\in\Omega^N}\Vert \mathbb{J}{L}_{1,\cdot}(S_1){L}_{1,x}(S_1)-\mathbb{J}{L}_{1,x}(S_1'){L}_{1,x}(S_1')\Vert\nn\\
&\le \Vert \mr{Ev}_{S_2}-\mr{Ev}_{S_2'}\Vert\sup_{x\in\Omega^N}\Vert \mathbb{J}{L}_{1,x}(S_1){L}_{1,x}(S_1)\Vert+\Vert \mr{Ev}_{S_2'}\Vert\sup_{x\in\Omega^N}\Vert \mathbb{J}{L}_{1,x}(S_1)\Vert \Vert{L}_{1,x}(S_1)-{L}_{1,x}(S_1')\Vert \nn\\
&\qquad +\Vert \mr{Ev}_{S_2'}\Vert\sup_{x\in\Omega^N}\Vert \mathbb{J}{L}_{1,x}(S_1)- \mathbb{J}{L}_{1,x}(S_1')\Vert\Vert{L}_{1,x}(S_1')\Vert\nn\\
&\le \epsilon\sup_{x\in\Omega^N}\Vert E_{S_1,x}\Vert\bigg\Vert \bigg[\Theta_{S_1}(\Phi(\Delta t,x))_i-\omega_{S_1}\Delta t-\Theta_{S_1}(x)-\sum_{j=1}^M\sum_{k=1}^N{a}_{S,i,k}^j \mr{e}^{\sqrt{-1}j(\Theta_{S_1}(x)_k-\Theta_{S_1}(x)_i)}\bigg]_i\bigg\Vert\nn\\
&\qquad +\Vert \mr{Ev}_{S_2'}\Vert\sup_{x\in\Omega^N}\Vert E_{S_1,x}\Vert \bigg(\Vert \Theta_{S_1}(\Phi(\Delta t,x))-\Theta_{S_1'}(\Phi(\Delta t,x))\Vert+\Delta t\vert\omega_{S_1}-\omega_{S_1'}\vert\nn\\
&\qquad+\sum_{j=1}^M\bigg\Vert \sum_{k=1}^N[({a}_{S,i,k}^j-{a}_{S',i,k}^j)\mr{e}^{\sqrt{-1}j(\Theta_{S_1}(x)_k-\Theta_{S_1}(x)_i)}]_i\bigg\Vert\nn\\
&\qquad+\sum_{j=1}^M\bigg\Vert \sum_{k=1}^N[{a}_{S',i,k}^j(\mr{e}^{\sqrt{-1}j(\Theta_{S_1}(x)_k-\Theta_{S_1}(x)_i)}-\mr{e}^{\sqrt{-1}j(\Theta_{S_1'}(x)_k-\Theta_{S_1'}(x)_i)})]_i\bigg\Vert\bigg)\nn\\
&\qquad+\Vert \mr{Ev}_{S_2'}\Vert\sup_{x\in\Omega^N}\Vert E_{S_1,x}-E_{S_1',x}\Vert(2\sqrt{N}+2\pi\Delta t+M\Vert A_{S'}^j\Vert\sqrt{N})\nn\\
&\le \epsilon N\sqrt{M}(2\sqrt{N}+2\pi\Delta t+M\Vert A_{S}^j\Vert\sqrt{N})\nn\\
&\qquad +\Vert \mr{Ev}_{S_2'}\Vert N\sqrt{M}(\epsilon+\Delta t\epsilon+MN\sqrt{N}\epsilon+M\Vert A_{S'}^j\Vert\sqrt{N}j\sup_{x\in\Omega^N}\vert (\Theta_{S_1}(x)_k-\Theta_{S_1}(x)_i)-(\Theta_{S_1'}(x)_k-\Theta_{S_1'}(x)_i)\vert)\nn\\
&\qquad +\Vert \mr{Ev}_{S_2'}\Vert NM\epsilon(2\sqrt{N}+2\pi\Delta t+M\Vert A_{S'}^j\Vert\sqrt{N})\nn\\
&\le \epsilon N\sqrt{M}(2\sqrt{N}+2\pi\Delta t+MD_A\sqrt{N}+D_E(1+\Delta t+MN\sqrt{N}+2M\sqrt{N}D_Aj))+\epsilon D_E N M(2\sqrt{N}+2\pi\Delta t+MD_A\sqrt{N})
\end{align*}
where $(E_x)_{i,j,k,i'}=\mr{e}^{\sqrt{-1}j(\Theta_{S}(x)_k-\Theta_{S}(x)_i)}$ if $i=i'$ and $(E_x)_{i,j,k,i'}=0$ if $i\neq i'$.
Thus, we have
\begin{align*}
\vert a^{\opn{new},j}_{S,i,k}-a^{\opn{new},j}_{S',i,k}\vert 
&\le\vert a^j_{S,i,k}-a^j_{S',i,k}\vert + \eta\vert \nabla L(S)-\nabla L(S')\vert_{i,j,k}\\
&\le \epsilon+2\epsilon \eta N\sqrt{M}(2\sqrt{N}+2\pi\Delta t+MD_A\sqrt{N}+D_E(1+\Delta t+MN\sqrt{N}+2M\sqrt{N}D_Aj))\\
&\quad+2\epsilon\eta D_E N M(2\sqrt{N}+2\pi\Delta t+MD_A\sqrt{N}).
\end{align*}

\end{proof}

\section{Experimental details}\label{ap:experiment}
\paragraph{Optimization}
For all experiments, to find the parameter $A^j=[a^j_{i,k}]_{i,k}$ that is close to the identity matrix, we set the initial value of $A^j$ as the identity matrix and added the regularization term $0.01\sum_{j=1}^Mj\Vert A^j-I\Vert_F^2$ for the KGME and Power approaches.
Here, $\Vert\cdot\Vert_F$ is the Frobenius norm.
The identity matrix is replaced by the zero matrix for the Fourier approach.
We restricted $a_{i,k}^j$ to the real value to obtain $\Gamma_i$ as an odd function.
In addition, to make the absolute value of the right-hand side of Eq.~\eqref{eq:phase_koopman} close to 1, we added the regularization term $0.01\sum_{x\in S_2}j\vert\vert\sum_{k=1}^Na_{i,k}^ju(x)\vert-1\vert^2$ for the KGME and Power approaches.
For Subsections~\ref{subsec:fhn} and \ref{subsec:mice}, we used the gradient descent method with a learning rate of 0.1 for optimizing $A^j$.
We applied 3000 iterations to obtain the results.

\paragraph{Function space}
For all experiments in this section, for the Hilbert spaces $\tilde{\hil}$ and $\hil$, we used the RKHS and vvRKHS associated with the Laplacian kernel $\tilde{k}(x,y)=\mr{e}^{-0.1\vert x-y\vert}$ and $k(x,y)=[\tilde{k}(x_i,y_j)]_{i,j}$ (see Appendix~\ref{ap:vvrkhs}).


\end{document}